\documentclass[amsmath,10pt,twocolumn,superscriptaddress,footinbib,pre]{revtex4-1}
\usepackage[utf8]{inputenc}
\usepackage{hyperref}
\usepackage{graphicx} 
\usepackage{amsmath,amsfonts,bm} 
\usepackage{uri}

\usepackage[x11names, rgb, html]{xcolor}
\definecolor{sbase03}{HTML}{002B36}
\definecolor{sbase02}{HTML}{073642}
\definecolor{sbase01}{HTML}{586E75}
\definecolor{sbase00}{HTML}{657B83}
\definecolor{sbase0}{HTML}{839496}
\definecolor{sbase1}{HTML}{93A1A1}
\definecolor{sbase2}{HTML}{EEE8D5}
\definecolor{sbase3}{HTML}{FDF6E3}
\definecolor{syellow}{HTML}{B58900}
\definecolor{sorange}{HTML}{CB4B16}
\definecolor{sred}{HTML}{DC322F}
\definecolor{smagenta}{HTML}{D33682}
\definecolor{sviolet}{HTML}{6C71C4}
\definecolor{sblue}{HTML}{268BD2}
\definecolor{scyan}{HTML}{2AA198}
\definecolor{sgreen}{HTML}{859900}
\hypersetup{
  colorlinks = true,
  citecolor = {sred},
  urlcolor = {sblue}
}

\newcommand{\tobs}{t_{\text{obs}}}
\newcommand{\blam}{\boldsymbol{\lambda}}

\newcommand{\bdeta}{\boldsymbol{\eta}}
\newcommand{\bxi}{\boldsymbol{\xi}}
\newcommand{\bzeta}{\boldsymbol{\zeta}}
\newcommand{\bzetat}{\tilde{\boldsymbol{\zeta}}}
\newcommand{\bxit}{\tilde{\boldsymbol{\xi}}}
\newcommand{\kAB}{k_{AB}}
\newcommand{\xvec}{\overrightarrow{\mathbf{x}}}
\newcommand{\xvecback}{\overleftarrow{\mathbf{x}}}
\newcommand{\xb}{\mathbf{x}}

\begin{document}
\title{Estimating Reciprocal Partition Functions to Enable Design Space Sampling}
\author{Alex Albaugh}
\author{Todd R.~Gingrich}
\affiliation{Department of Chemistry, Northwestern University, 2145 Sheridan Road, Evanston, Illinois 60208, USA}

\begin{abstract}
Reaction rates are a complicated function of molecular interactions, which can be selected from vast chemical design spaces.
Seeking the design that optimizes a rate is a particularly challenging problem since the rate calculation for any one design is itself a difficult computation.
Toward this end, we demonstrate a strategy based on transition path sampling to generate an ensemble of designs and reactive trajectories with a preference for fast reaction rates.
Each step of the Monte Carlo procedure requires a measure of how a design constrains molecular configurations, expressed via the reciprocal of the partition function for the design.
Though the reciprocal of the partition function would be prohibitively expensive to compute, we apply Booth's method for generating unbiased estimates of a reciprocal of an integral to sample designs without bias.
A generalization with multiple trajectories introduces a stronger preference for fast rates, pushing the sampled designs closer to the optimal design.
We illustrate the methodology on two toy models of increasing complexity: escape of a single particle from a Lennard-Jones potential well of tunable depth and escape from a metastable tetrahedral cluster with tunable pair potentials.
\end{abstract}
\maketitle

\section{Introduction}
\label{sec:intro}
One of the most challenging and important aims of theoretical and computational chemistry is the calculation of rates. 
The speed of chemical events can vary over many orders of magnitude, ranging from electron transfer on a femtosecond timescale to material aging over millennia.
Direct simulation of quantum or classical dynamics can provide access to these fastest timescales, but numerically computing rates for activated processes is notoriously difficult due to the rare event problem~\cite{kramers1940brownian, hanggi1990reaction, bolhuis2002transition,peters2010recent}.
To combat this problem, several related methodologies have been developed based on the connection between time correlation functions and rate constants~\cite{yamamoto1960quantum,miller1983quantum,chandler1978statistical,montgomery1979trajectory,chandler1986roles,chandler1988two,borkovec1990generalized,bolhuis2002transition,hummer2004transition}.
Crucially, those correlation functions can often be calculated from dynamical trajectories of modest length using methods like transition path sampling~\cite{pratt1986statistical,bolhuis2002transition,dellago2002transition,bolhuis2002transition2,dellago1998transition,bolhuis1998sampling,dellago1999calculation,peters2006obtaining,peters2007extensions,miller2007sampling,grunwald2008precision}, 
transition interface sampling~\cite{van2003novel,moroni2004rate,van2005elaborating}, and forward flux sampling~\cite{allen2005sampling,allen2006simulating,allen2006forward,allen2009forward,borrero2008optimizing,borrero2009simulating}.

In principle, one should thus be able to attack chemical design problems—problems like determining what side chains of a peptide most effectively amplify a rate of catalysis.
In practice, it becomes very expensive to perform a converged rate calculation for every candidate design.
One approach to circumvent this expense is to \emph{sample} the design space with a random walker, statistically biased to spend most of its time visiting designs with fast rates.
Let \(\blam\) denote all parameters one seeks to design; these could be particle charges, amino acid identities, Lennard-Jones parameters, bond strengths, equilibrium bond lengths, etc.
Each design has some rate constant \(k(\blam)\), and one might hope to sample possible designs from the probability distribution with probability density \(P(\blam) \propto k(\blam)\), thereby giving extra statistical weight to those designs with faster rates. 
A simple, straightforward way to sample designs is to carry out a Markov Chain Monte Carlo (MCMC) simulation, consisting of three iterated steps:
(1) attempt to transition from design \(\blam\) to a new design \(\blam'\) with probability \(P_{\rm gen}(\blam \to \blam')\);
(2) compute a converged rate calculation of both \(k(\blam)\) and \(k(\blam')\);
(3) accept the new design with probability
\begin{equation}
P_{\rm acc}[\blam \to \blam'] = \min\left[1, \frac{P_{\rm gen}(\blam' \to \blam) k(\blam')}{P_{\rm gen}(\blam \to \blam') k(\blam)}\right].
\label{eq:designMC}
\end{equation}
While this procedure would steer the sampled designs toward those with faster rates, it requires high-quality converged rate calculations for every proposed \(\blam\).

To radically reduce the computational expense, one might instead hope to carry out the MCMC dynamics with noisy estimates for \(k(\blam)\), akin to Ceperley and Dewing's penalty method for random walks with noisy energies~\cite{ceperley1999penalty}.
The essential idea is to execute a random walk in the higher dimensional space of designs \emph{and} reactive trajectories, those that transition from reactant to product in a fixed observation time \(\tobs\).
Every step of the Monte Carlo procedure outlined in Eq.~\eqref{eq:designMC} requires a converged rate calculation to decide whether to accept a newly proposed design, but each step of the joint-space random walker uses a noisy estimate of that rate.
This noisy estimate characterizes how probable it is to generate a reactive trajectory given the design \(\blam\), assuming the trajectory was initialized in an equilibrium reactant configuration.

The requirement that reactive trajectories be initialized in an equilibrium ensemble presents a significant technical problem, the resolution of which is the focus of this manuscript.
The challenge is that computing the acceptance probability for a Monte Carlo step requires the Boltzmann probability of the initial condition, which depends on a design-dependent canonical partition function.
Were the design held fixed, a ratio of identical partition functions would cancel in the Monte Carlo acceptance formulae.
Without that cancellation, the MCMC procedure requires unbiased estimates of the reciprocal of the partition function.
In this manuscript, we show how those estimates can be obtained using Booth's method for generating unbiased estimates of integrals~\cite{booth2007unbiased}.
By applying that strategy to the rate design problem, we can sample \( P(\blam) \propto k(\blam) \) without converged rate estimates.
A stronger preference for designs with large rates is applied by introducing \(L\) independent trajectories to sample \(P(\blam) \propto k(\blam)^L\).

This manuscript is split into two parts.
We review and develop the theoretical tools in Section~\ref{sec:theory}.
We then illustrate the methodology on two model systems in Section~\ref{sec:results}.
The first simple system---sampling different strengths of attraction between two Lennard-Jones particles in proportion to their unbinding rate---serves to validate the mathematics by comparing results from sampling with explicit numerical calculations.
The second system---sampling a three-dimensional design space of interaction parameters between a particle metastably trapped in a cage---more clearly illustrates the potential benefits when studying more complex, higher-dimensional problems.

\section{Theory}
\label{sec:theory}
\subsection{Reaction rates and trajectory-space sampling}
\label{sec:rates}
Consider the phase space of a classical system, \( \mathbf{x}=\{\mathbf{r},\mathbf{p}\} \), which consists of a set of positions of all \(N\) particles, \( \mathbf{r} = \{\mathbf{r}_{1},\mathbf{r}_{2} \ldots\mathbf{r}_{N}\} \), and their momenta, \( \mathbf{p} = \{\mathbf{p}_{1},\mathbf{p}_{2} \ldots\mathbf{p}_{N}\} \).  
The total energy of this system is given by the Hamiltonian \(H\), the sum of the kinetic energy \(K\) and potential energy \(U\): \(H(\mathbf{x};\blam) = U(\mathbf{r};\blam)+K(\mathbf{p};\blam)\).
The energy depends not only on \( \mathbf{r} \) but also on some parameters \(\blam\), which could include particle charges, Lennard-Jones parameters, bond strengths, etc.
In this work we imagine these ``design parameters'' to be time-independent and controllable .
A system evolving under \(H(\mathbf{x};\blam)\) traces out trajectories in phase space that we denote \( \overrightarrow{\mathbf{x}} \).
We will focus on discrete time-evolution generated by numerical integration such that the trajectory is a sequence of \( M + 1 \) points in phase space separated by increments of time \( \Delta t \): \( \overrightarrow{\mathbf{x}} = \{ \mathbf{x}(0), \mathbf{x}(\Delta t), \hdots, \mathbf{x}(\tobs)\} \), with observation time \(\tobs = M \Delta t\).

Chemical systems tend to be high-dimensional with potential energy surfaces that possess multiple metastable basins.
Trajectories occupy a metastable region of phase space for relatively long periods of time before making rare transitions to another metastable region.
In the simplest scenario, there are two principal, non-overlapping basins, \(A\) and \(B\), which correspond respectively to reactants and products.
Provided the transitions are rare, there exists a first-order rate constant \(\kAB(\blam)\) that depends on the particular design.
If each \(A \to B\) transition is independent of previous transitions, e.g., if memory is lost, then the process is Poissonian with the time between reactions, \(\tau\), coming from the distribution
\begin{equation}
P(\tau|\blam) = \kAB(\blam) e^{-\kAB(\blam) \tau}.
\label{eq:Poissonwaiting}
\end{equation}
The probability that a trajectory, starting in \(A\), will exhibit at least one reaction in time \(\tobs\) is thus given by
\begin{align}
\nonumber
\int_0^{\tobs} d\tau \, \kAB(\blam) e^{- \kAB(\blam) \tau} &= 1 - e^{-\kAB(\blam) \tobs}\\
&\approx \kAB(\blam) \tobs.
\label{eq:poisson}
\end{align}
The final approximation is justified when \(\tobs \ll 1 / \kAB(\blam)\), in which case trajectories only have time for either zero or one reaction event.

Following Ref.~\cite{bolhuis2002transition}, the probability of a reaction can also be computed as
\begin{align}
\nonumber &P_{\text{reaction}}(\blam, \tobs)\\
& \ \ = \frac{\int \mathcal{D}\xvec \, h_A(\xb(0)) h_B(\xb(\tobs)) e^{-\beta H(\xb(0);\blam)} P(\xvec | \blam, \xb(0))}{Z_A(\blam)},
\label{eq:reactive2}
\end{align}
where \(\beta = (k_{\rm B} T)^{-1}\) is the inverse temperature, \(k_{\rm B}\) is Boltzmann's constant, and\(h_A\) and \(h_B\) are indicator functions that evaluates to zero or one so as to constrain trajectories to end as products.
Here the probability of trajectory \(\xvec\) has been decomposed into a product of the equilibrium Boltzmann probability for the initial reactant configuration \(\xb(0)\) times the (normalized) probability of subsequent dynamics given that initialization, \(P(\xvec | \blam, \xb(0))\).
The canonical partition function of the reactant state, 
\begin{equation}
Z_A(\blam) = \int d\mathbf{x}(0) \, h_A(\mathbf{x}(0)) e^{-\beta H(\xb(0); \blam)},
\end{equation}
measures how \(\blam\) impacts the Boltzmann probability for being initialized in a reactant configuration.

Provided \(\tobs \ll 1 / \kAB(\blam)\), the Poisson description of Eq.~\eqref{eq:poisson} can be equated with the trajectory-space average of Eq.~\eqref{eq:reactive2}.
That equality expresses the rate constant in terms of a trajectory space average,
\begin{equation}
\kAB(\blam) = \frac{1}{\tobs} \int \mathcal{D}\xvec \, \rho(\xvec, \blam),
\label{eq:kabint}
\end{equation} 
where
\begin{equation}
\rho(\xvec, \blam) = h_A(\xb(0)) h_B(\xb(\tobs)) \frac{e^{-\beta H(\xb(0);\blam)}}{Z_A(\blam)} P(\xvec|\blam, \xb(0)).
\label{eq:rho}
\end{equation}
Upon normalizing, \(P(\xvec, \blam) = \rho(\xvec, \blam) / \mathcal{N}\) has the interpretation of a joint distribution over trajectories and designs, with normalization
\begin{equation}
\mathcal{N} = \int d\blam \, \int \mathcal{D}\xvec \, \rho(\xvec, \blam).
\end{equation}
If one samples trajectories and designs from that distribution, the marginal distribution over designs would therefore be
\begin{equation}
P(\blam) = \int \mathcal{D} \xvec \, P(\xvec, \blam) = \frac{\tobs}{\mathcal{N}} \kAB(\blam).
\label{eq:problam}
\end{equation}
The ratio of rates for designs \(\blam\) and \(\blam'\) is consequently the relative likelihood of observing \(\blam\) and \(\blam'\) in the sampling of \(P(\xvec, \blam)\):
\begin{equation}
\frac{P(\blam)}{P(\blam')} = \frac{\kAB(\blam)}{\kAB(\blam')}.
\label{eq:ratio}
\end{equation}

\begin{figure*}[ht]
\includegraphics[width=0.9\textwidth]{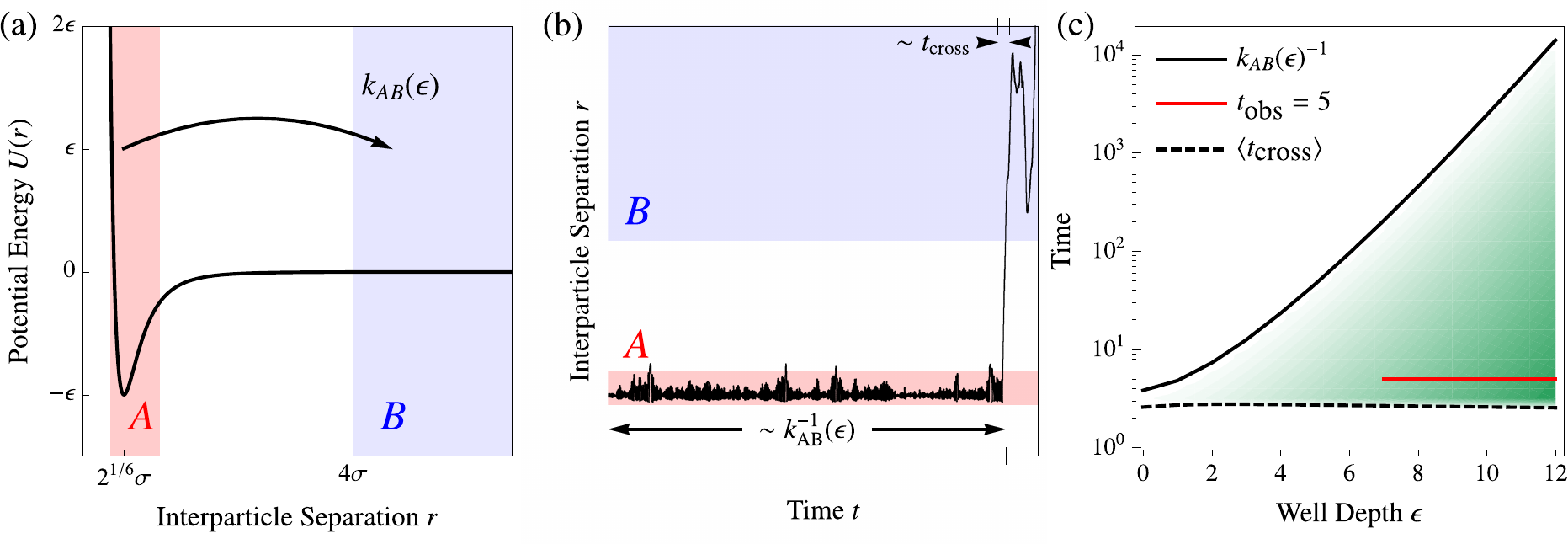}
\caption{
(a) Rate constants can be calculated by computing the probability of rare trajectories that execute transitions between states.
Those rate constants depend on the molecular design.
In the test system of Section~\ref{sec:results} the molecular design is the choice of well depth of a Lennard-Jones potential.
The rare reactive trajectories start in region \(A\) and end in region \(B\), given by the shaded red and blue areas, respectively.
(b) An example of an unbinding trajectory that escapes from \(A\) to \(B\).
The timescale of the overall rate is labeled as \(\kAB^{-1}\) and the much shorter timescale of just the escape event is labeled as \(t_{\text{cross}}\).
(c) The transition path sampling framework for computing rates requires the separation of timescales: \(t_{\text{cross}} < \tobs \ll \kAB^{-1}\).
The possible Lennard-Jones well depth \(\epsilon\) and observation time \(\tobs\) must be selected so the sampling is confined to the green shaded region, where that timescale separation is valid.
The red line represents the \(\tobs\) and \(\epsilon\) range that we numerically sampled in Sections~\ref{sec:traj} and~\ref{sec:multi}.
}
\label{fig:Fig1}
\end{figure*}

The goal of sampling \(P(\blam) \propto \kAB(\blam)\) has been reduced to the problem of sampling the joint distribution \(P(\xvec, \blam)\) provided \(\tobs \ll 1 / \kAB\).
Eq.~\eqref{eq:ratio} would break down if \(\tobs\) were too large, so one may imagine using arbitrarily small \(\tobs\).
That choice results in a different issue.
Rare \(A \to B\) transitions occupy \(A\) for a comparatively long time before carrying out a rapid passage over the barrier.
By pushing to smaller \(\tobs\), one can excise some of that waiting time without impacting the mechanism of the barrier crossing, but there is a minimum amount of time, \(t_{\rm cross}\), needed to cross.
Using an observation time that is less than this crossing time also causes Eq.~\eqref{eq:ratio} to break down.
We therefore require that a suitable \(\tobs\) is chosen such that both timescale restrictions are met \((t_{\rm cross} < \tobs \ll 1 / \kAB)\) for all sampled designs.
The necessary timescale separation is illustrated explicitly for our Lennard-Jones unbinding problem in Fig.~\ref{fig:Fig1}.

\subsection{The reciprocal partition function problem}
\label{sec:recZ}
The strategy of Section~\ref{sec:rates} allowed us to convert the problem of sampling \( P(\blam) \propto \kAB(\blam) \) into the higher-dimensional joint problem \(P(\xvec, \blam) \propto \rho(\xvec, \blam),\) with \(\rho\) given by Eq.~\eqref{eq:rho}.
That higher-dimensional space can be sampled with a Metropolis-Hastings MCMC procedure by proposing a change from \(\xvec, \blam\) to some new \(\xvec', \blam'\) according to a generation probability \(P_{\rm gen}(\xvec, \blam \to \xvec', \blam')\).
That proposal move is then conditionally accepted with probability 
\begin{equation}
P_{\rm acc} = \min\left[1, \frac{\rho(\xvec', \blam') P_{\rm gen}(\xvec', \blam' \to \xvec, \blam)}{\rho(\xvec, \blam) P_{\rm gen}(\xvec, \blam \to \xvec', \blam')}\right].
\label{eq:pacc}
\end{equation}
The acceptance probability depends on the manner that new designs and trajectories are generated, that is on \(P_{\rm gen}\).
Specific choices of proposal moves are discussed in Appendix~\ref{sec:trajapp}, but a typical feature of those strategies is that the ratio of generation probabilities can be explicitly computed for any \(\xvec, \blam\) and \(\xvec', \blam'\).
In contrast, it is \emph{not} typically possible to compute the ratio of \(\rho\) factors in Eq.~\eqref{eq:pacc}:
\begin{align}
\nonumber
\frac{\rho(\xvec', \blam')}{\rho(\xvec, \blam)} &= \frac{Z_A(\blam')^{-1}}{Z_A(\blam)^{-1}} \times \frac{e^{-\beta H(\xb'(0);\blam')}}{e^{-\beta H(\xb(0);\blam)}}\\
 & \ \ \  \times \frac{h_A(\xb'(0)) h_B(\xb'(\tobs))}{h_A(\xb(0)) h_B(\xb(\tobs))} \times \frac{P(\xvec'|\blam', \xb'(0))}{P(\xvec|\blam, \xb(0))}.
\label{eq:rhoratio}
\end{align}
The only problematic term is the ratio of the reciprocal of the partition functions.
Upon proposing a new \(\blam'\), Eqs.~\eqref{eq:pacc} and~\eqref{eq:rhoratio} require that one compute \(Z_A(\blam')^{-1}\), but computing a partition function is computationally expensive.
Even if one were to exhaustively compute \(Z_A(\blam')\) by sampling phase space, the partition function would only be known up to some sampling error, and an unbiased estimate for \(Z_A\) would give a biased estimate for \(Z_A^{-1}\).
Inserting that biased noise into the acceptance probability would bias the Markov chain's stationary distribution.

The problem is quite similar to Ceperley and Dewing's consideration of Monte Carlo with noisy energies~\cite{ceperley1999penalty}, except now the noise comes from imperfect computations of the \(Z_A(\blam)^{-1}\) terms.
The resolution is to replace \(Z_A(\blam)^{-1}\) by an unbiased estimate \(\widehat{Z_A(\blam)^{-1}}(\bdeta)\), where the variables \(\bdeta\) are all of the random numbers drawn from a distribution \(P(\bdeta)\) and used to estimate the reciprocal partition function.
For example, if the estimate requires one to compute energies of representatively sampled configurations, \(\bdeta\) would be the random numbers necessary to construct such samples and \(P(\bdeta)\) would be built up from the Gaussian or uniform distributions that the computer's random number generator used to select those random numbers.
The unbiased estimate will appear naturally in the acceptance probability when one samples \(\xvec, \blam,\) and \(\bdeta\) in proportion to
\begin{align}
\nonumber \tilde{\rho}(\xvec, \blam, \bdeta) &= P(\bdeta) \widehat{Z_A(\blam)^{-1}}(\bdeta) h_A(\xb(0)) h_B(\xb(\tobs)) \\
& \ \ \ \ \ \ \ \ \ \ \ \times e^{-\beta H(\xb(0);\blam)} P(\xvec|\blam, \xb(0)).
\label{eq:lift}
\end{align}

To be explicit, proposed changes \(\xvec, \blam, \bdeta \to \xvec', \blam', \bdeta'\) are accepted with probability
\begin{equation}
P_{\rm acc} = \min\left[1, \frac{\tilde{\rho}(\xvec', \blam', \bdeta') P_{\rm gen}(\xvec', \blam' \to \xvec, \blam) P_{\rm gen}(\bdeta)}{\tilde{\rho}(\xvec, \blam, \bdeta) P_{\rm gen}(\xvec, \blam \to \xvec', \blam') P_{\rm gen}(\bdeta')}\right].
\label{eq:pacc2}
\end{equation}
We assume that the new estimate of the reciprocal partition function is generated by drawing new random numbers from \(P_{\rm gen}(\bdeta') = P(\bdeta')\).
As before, the ratio of \(P_{\rm gen}\) terms for \(\xvec\) and \(\blam\) in Eq.~\eqref{eq:pacc} can be explicitly computed.
The remaining ratio takes the same form as Eq.~\eqref{eq:rhoratio} except that \(Z_A(\blam)^{-1}\) has been replaced by the estimate \(\widehat{Z_A(\blam)^{-1}}\):
\begin{align}
\nonumber
&\frac{\tilde{\rho}(\xvec', \blam', \bdeta') P_{\rm gen}(\bdeta)}{\tilde{\rho}(\xvec, \blam, \bdeta)P_{\rm gen}(\bdeta')} = \frac{\widehat{Z_A(\blam')^{-1}}(\bdeta')}{\widehat{Z_A(\blam)^{-1}}(\bdeta)} \times \frac{e^{-\beta H(\xb'(0);\blam')}}{e^{-\beta H(\xb(0);\blam)}} \\
 & \ \ \ \ \ \ \ \ \ \times \frac{h_A(\xb'(0)) h_B(\xb'(\tobs))}{h_A(\xb(0)) h_B(\xb(\tobs))} \times \frac{P(\xvec'|\blam', \xb'(0))}{P(\xvec|\blam, \xb(0))}.
\label{eq:rhoratio2}
\end{align}
The resulting MCMC procedure in \(\xvec, \blam, \bdeta\) space can therefore accept and reject proposal moves based on the noisy estimate in lieu of the intractable reciprocal partition function.
It is important to note that since the reciprocal partition function estimates is computed from \(\bdeta\) which, upon rejection, does not get updated to the trial value \(\bdeta'\).
As such, the old estimate of the reciprocal partition function with \(\bdeta\) must be ``reused'' until the next move is accepted.

By choosing a noisy estimate that is unbiased, we ensure that we will recover the original \(\rho\) after marginalizing over the \(\bdeta\) variables:
\begin{equation}
\rho(\xvec, \blam) = \int d\bdeta \, \tilde{\rho}(\xvec, \blam, \bdeta)
\end{equation}
because
\begin{equation}
\left<\widehat{Z_A(\blam)^{-1}}\right> = \int d\bdeta \, P(\bdeta) \widehat{Z_A(\blam)^{-1}}(\bdeta) = Z_A(\blam)^{-1}.
\end{equation}
Indeed, this marginalization step required that we expressed Eq.~\eqref{eq:lift} in terms of an estimate of the reciprocal of the partition function, not a reciprocal of an estimate of the partition function.
The latter would have been simpler to estimate, but it would not have marginalized to give Eq.~\eqref{eq:rho}.
Notably, we have not required that the acceptance probabilities computed in Eq.~\eqref{eq:pacc2} are unbiased estimates for those of Eq.~\eqref{eq:pacc}; they generally will not be.

The sampling scheme does not assume a low-variance estimator, only one which is unbiased, but that variance can affect computational performance.
A large-variance estimate typically results in more Monte Carlo rejections than one with low variance~\cite{ceperley1999penalty, gingrich2015preserving}, but it could nevertheless be advantageous to use the large-variance estimate if it is particularly cheap to compute.

From one perspective, the strategies employed are an exercise in the usefulness of lifts to Monte Carlo methods~\cite{diaconis2000analysis,lin2000noisy,andrieu2009pseudo,bernard2009event,vucelja2016lifting}.
We ultimately are interested in sampling \(P(\blam)\), a distribution over possible designs, but we access that distribution by targeting higher-dimensional distributions.
A lift from \(\blam\) to \(\left(\xvec, \blam\right)\) left us with the problematic reciprocal partition function, which we subsequently replaced by an estimate via a second lift to \(\left(\xvec, \blam, \bdeta\right)\).
To utilize this final lift, we require a method for generating unbiased estimates of the reciprocal partition function, a problem addressed in a more general setting by Booth~\cite{booth2007unbiased}.

\subsection{Estimating reciprocals of partition functions}
\label{sec:recipest}

The partition function \(Z_A(\blam)\) involves an integral over all of phase space, both \(\mathbf{r}\) and \(\mathbf{p}\).
Because the Hamiltonian decouples into a potential energy depending on positions and a kinetic energy depending on momenta, the (classical) partition function can be decomposed as
\begin{equation}
Z_A(\blam) = \frac{1}{C} \bar{Z}(\blam) \tilde{Z}_A(\blam),
\end{equation}
where 
\begin{equation}
\tilde{Z}_A(\blam) = \int d\mathbf{r} \, h_A(\mathbf{r}) e^{-\beta U(\mathbf{r};\blam)}
\end{equation}
is the configurational partition function,
\begin{equation}
\bar{Z}(\blam) = \int d\mathbf{p} \, e^{-\beta K(\mathbf{p}; \blam)}.
\end{equation}
is the partition function for the momenta, and \(C\) is a constant that handles the exchange symmetry for identical particles and the discretization of phase space.
For example, the case of identical classical particles in three dimensions gives \(C = h^{3N} N!\), where \(h\) is Planck's constant.
The integral over momenta \(\bar{Z}(\blam)\) does not need to be estimated because the quadratic form of kinetic energy allows it to be computed explicitly as a Gaussian integral.
In contrast, for all but the simplest potential energies, we must estimate the configurational contribution to get estimates of the reciprocal partition functions:
\begin{equation}
\widehat{Z_A(\blam)^{-1}} = C \bar{Z}(\blam)^{-1} \widehat{\tilde{Z}_A(\blam)^{-1}}.
\end{equation}
In this work we limit ourselves to changes of design that alter neither \(C\) nor \(\bar{Z}\), in which case the contribution to a Monte Carlo acceptance ratio comes from the \(\widehat{\tilde{Z}_A(\blam)^{-1}}\) term.

That term is the reciprocal of an integral over \(\mathbf{r}\), precisely the situation where Booth's method provides an unbiased estimate~\cite{booth2007unbiased}.
The core insight behind Booth's approach is to replace the reciprocal of \(\tilde{Z}_A(\blam)\) by a series expansion and to generate unbiased estimates for each term in that series.  
A similar idea, extended to partition functions of general probability distributions, is outlined in~\cite{lyne2015russian}.
Using a geometric series, the expansion can be written in terms of a fixed reference design \(\blam_{\rm ref}\) as
\begin{align}
\nonumber \frac{1}{\tilde{Z}_{A}(\blam)} &= \frac{1}{\int d\mathbf{r} \ h_{A}(\mathbf{r}) \, e^{-\beta U(\mathbf{r}; \blam)}}\\
&= \frac{\left(1 + a \left(1 + a \left(1 + a \left(1 + \hdots\right)\right)\right)\right)}{\int d\mathbf{r} \ h_{A}(\mathbf{r}) \, e^{-\beta U(\mathbf{r}; \blam_{\rm ref})}} ,
\label{eq:series}
\end{align}
where
\begin{equation}
a = \frac{\int d\mathbf{r} \ h_{A}(\mathbf{r}) \ \left(e^{-\beta U(\mathbf{r};\blam_{\rm ref})} - e^{-\beta U(\mathbf{r};\blam)}\right)}{\int d\mathbf{r} \ h_{A}(\mathbf{r}) \ e^{-\beta U(\mathbf{r};\blam_{\rm ref})}} = 1 - \frac{\tilde{Z}_A(\blam)}{\tilde{Z}_A(\blam_{\rm ref})}
\end{equation}
must have a modulus less than one for the series to converge.
Though \(a\) depends on \(\blam\), we have suppressed that dependence in the notation.
It is generally intractable to compute \(a\) exactly, but we can get unbiased estimates by sampling independent configurations \(\mathbf{r}^{(1)}, \mathbf{r}^{(2)}, \hdots \) from the reference Boltzmann distribution \(P(\mathbf{r};\blam_{\rm ref}) =  h_{A}(\mathbf{r}) \ e^{-\beta U(\mathbf{r};\blam_{\rm ref})}/\tilde{Z}_A(\blam_{\rm ref})\).
The \(i^{\rm th}\) sampled configuration corresponds to the \(i^{\rm th}\) estimate
\begin{equation}
a^{(i)} = 1 - e^{-\beta (U(\mathbf{r}^{(i)};\blam) -  U(\mathbf{r}^{(i)};\blam_{\rm ref}) ) }.
\label{eq:a}
\end{equation}
It is straightforward to confirm that each estimate \(a^{(1)}, a^{(2)}, \hdots\) is unbiased:
\begin{align}
\nonumber \langle a^{(i)} \rangle &= \int d\mathbf{r}^{(i)} \, a^{(i)} P(\mathbf{r}^{(i)};\blam_{\rm ref})\\
&= 1 - \frac{ \int d\mathbf{r}^{(i)} h_{A}(\mathbf{r}) e^{-\beta U(\mathbf{r}^{(i)};\blam)} }{\tilde{Z}_A(\blam_{\rm ref})}  = a.
\end{align}
It follows that an unbiased estimate for \(\tilde{Z}(\blam)^{-1}\) can be constructed by replacing each instance of \(a\) in Eq.~\eqref{eq:series} by \(a^{(1)}, a^{(2)},\) etc.:
\begin{equation}
\widehat{\frac{1}{\tilde{Z}_{A}(\blam)}} = \frac{1}{\tilde{Z}_A(\blam_{\rm ref})} \left(1 + a^{(1)} \left(1 + a^{(2)} \left(1 + \hdots\right)\right)\right).
\label{eq:estseries}
\end{equation}

As written, the series would require \(a\) to be estimated an infinite number of times.
To be practically useful, we must convert the infinite series into a finite sum, ideally one with few terms.
Truncation of the series at finite order, however, would introduce a bias to the estimate.
Like Bhanot and Kennedy's unbiased estimates of \(e^x\)~\cite{bhanot1985bosonic,lin2000noisy}, Booth constructed an unbiased stochastic truncation from a ``roulette procedure'' that randomly chooses when to truncate the series in Eq.~\eqref{eq:estseries}~\cite{booth2007unbiased}.
We use a similar roulette procedure where samples \(a^{(1)}, a^{(2)}, \hdots\) are generated one by one.
After sample \(a^{(n)}\) is generated, it is either incorporated into the nested product or it triggers the termination.
Whether to incorporate \(a^{(n)}\) is determined by two factors: a tunable parameter \(0 < R < 1\) and the running product
\begin{equation}
\Pi^{(n)} \equiv \left|a^{(n)}\right| \prod_{i=1}^{n-1} \left|a^{(i)}\right| \max\left[1, \frac{R}{\Pi^{(i)}}\right].
\label{eq:runningproduct}
\end{equation}
If \(\Pi^{(n)} < R\), then the series is truncated with probability \(1 - (\Pi^{(n)}/R)\).
In the event of truncation, the estimate is constructed from the first \(n-1\) terms as
\begin{equation}
\widehat{\frac{1}{\tilde{Z}_{A}(\blam)}} = \frac{1}{\tilde{Z}_A(\blam_{\rm ref})} \left(1 + \sum_{i=1}^{n-1} \prod_{j=1}^i a^{(j)} \max\left[1, \frac{R}{\Pi^{(j)}}\right]\right).
\label{eq:truncseries}
\end{equation}
Appendix~\ref{sec:unbiased} explicitly shows that this stochastic truncation yields an unbiased estimate for the reciprocal of the partition function.

\subsection{Sampling configurations from the reference distribution}
\label{sec:refdist}
In practice, our unbiased estimates \(a^{(i)}\) come from a library of pre-computed configurations \(\mathbf{r}^{(i)}\), drawn from samples of the Boltzmann distribution \(P(\mathrm{r}; \blam_{\rm ref}) \propto e^{-\beta U(\mathbf{r}; \blam_{\rm ref})}\).  
Using a standard canonical sampling procedure with a fixed reference \(\blam_{\rm ref}\), we store \(K\) independent configurations \(\mathbf{r}^{(i)}\) along with their respective reference potential energies \(U(\mathbf{r}^{(i)};\blam_{\rm ref})\).  
These \(K\) samples comprise a library that is generated only once.
Every estimate of \(a\) is generated by drawing a configuration uniformly from this library and calculating \(a^{(i)}\) from Eq.~\eqref{eq:a}.
Hence \(a\) can be estimated for each new value of \(\blam\) using the same pre-sampled reference states.

The effectiveness of the method depends critically on the choice of the reference parameters \(\blam_{\text{ref}}\).  
Recall that for the series of Eqs.~\eqref{eq:series} and~\eqref{eq:estseries} to converge, we assumed \(|a^{(i)}(\blam)| < 1\) for all \(i\), an assumption that is guaranteed by choosing a reference with
\begin{equation}
U(\mathbf{r}; \blam_{\text{ref}}) < U(\mathbf{r}; \blam) + \frac{\ln{2}}{\beta}
\label{eq:lambdarefineq}
\end{equation}
for each sampled configuration \(\mathbf{r}\).
Since the series should converge throughout the design sampling process, we furthermore want Eq.~\eqref{eq:lambdarefineq} to hold for all designs \(\blam\).
The reference energy can be made sufficiently low in two ways.
First, we can seek as \(\blam_{\text{ref}}\) the parameters \(\blam\) that minimize \(U(\mathbf{r};\blam)\) for all configurations \(\mathbf{r}\), but a globally optimal \(\blam_{\text{ref}}\) may not exist.
Indeed, if the minimizing \(\blam\) depends on the particular configuration \(\mathbf{r}\), it is necessary to sample configurations according to a shifted reference energy \(U(\mathbf{r}; \blam_{\text{ref}}) + U_0\).
In that case, \(\blam_{\text{ref}}\) could be any \(\blam\) (low energy is better) and the constant offset \(U_0\) is chosen such that
\begin{equation}
U_0  < \min_{\mathbf{r}} \left(U(\mathbf{r};\blam) - U(\mathbf{r}; \blam_{\text{ref}})\right) + \frac{\ln 2}{\beta}.
\label{eq:U0shift}
\end{equation}

These conditions on the reference energy ensure series convergence, but the guarantee comes with a computational cost.
By shifting to a lower offset energy, the series truncates after more terms, a trend that is easily rationalized in the \(U_0 \to -\infty\) limit.
Then every \(a^{(n)}\) tends to \(1\), so \(\Pi^{(n)}\) is very slow to decay below \(R\).
Consequently, the series seldom chooses to terminate.
A rapidly truncated convergent series thus demands a reference energy that is as high as possible without ever violating Eq.~\eqref{eq:U0shift}.

\subsection{Biasing for faster rates with multiple trajectories}
\label{sec:multitraj}
Section~\ref{sec:rates} illustrated how to sample in proportion to a transition rate \(\kAB(\blam)\).
Suppose, however, that the vast design space has a large design entropy.
The many designs with slow rates would overwhelm the probability of sampling one of the comparatively few designs with fast rates.
For the sampling procedure to discover those designs with anomalously fast transition rates, it generally requires a stronger bias in favor of fast rates.
For example, one could sample designs in proportion to \(\kAB(\blam)^L\) for some \(L\) greater than one.
If \(L\) is an integer, this more strongly biased distribution can be sampled in analogy with Section~\ref{sec:rates} by making use of \(L\) independent reactive trajectories~\cite{gingrich2016near}, collectively sampling the distribution
\begin{align}
P(\xvec_{1}, \xvec_{2}, \dots, \xvec_{L}, \blam) &= \frac{1}{\mathcal{N}^{L}}\prod_{l=1}^{L} h_A(\xb_{l}(0)) h_B(\xb_{l}(\tobs))
\\& \times \  \frac{e^{-\beta H(\xb_{l}(0);\blam)}}{Z_{A}(\blam)} P(\xvec_{l}|\blam, \xb_{l}(0)).
\label{eq:jointtrajectory}
\end{align}
Assuming the same timescale separation that led to Eq.~\eqref{eq:problam}, integration over the trajectories indeed leaves the targeted marginal distribution
\begin{align}
\nonumber P(\blam) &= \int \mathcal{D} \xvec_{1} \int \mathcal{D} \xvec_{2} \dots \int \mathcal{D} \xvec_{L}  \, P(\xvec_{1}, \xvec_{2}, \dots, \xvec_{L}, \blam) 
\\ &= \left[ \frac{\tobs}{\mathcal{N}} \kAB(\blam) \right]^{L}.
\label{eq:problammulti}
\end{align}

The multiple-trajectory joint distribution of Eq.~\eqref{eq:jointtrajectory} can be sampled with Metropolis-Hastings MCMC by proposing changes from \((\xvec_{1}, \xvec_{2}, \dots, \xvec_{L}, \blam) \to (\xvec_{1}', \xvec_{2}', \dots, \xvec_{L}', \blam')\) according to some calculable generation probability \(P_{\rm gen}\).  The changes are conditionally accepted with acceptance probability
\begin{align}
\nonumber P_{\rm acc} &= \min \left[1, \frac{P(\xvec_{1}', \xvec_{2}', \dots, \xvec_{L}', \blam')}{P(\xvec_{1}, \xvec_{2}, \dots, \xvec_{L}, \blam)}  \right.
\\& \times \left. \frac{P_{\rm gen}(\xvec_{1}', \xvec_{2}', \dots, \xvec_{L}', \blam' \to \xvec_{1}, \xvec_{2}, \dots, \xvec_{L}, \blam)}{ P_{\rm gen}(\xvec_{1}, \xvec_{2}, \dots, \xvec_{L}, \blam \to \xvec_{1}', \xvec_{2}', \dots, \xvec_{L}', \blam')} \right].
\label{eq:paccmulti}
\end{align}

Similar to the case of a single trajectory, one finds that the ratio of probabilities contains problematic ratios of reciprocal partition functions:
\begin{align}
\nonumber \frac{P(\xvec_{1}', \xvec_{2}', \dots, \xvec_{L}', \blam')}{P(\xvec_{1}, \xvec_{2}, \dots, \xvec_{L}, \blam)} = \frac{Z_{A}(\blam')^{-L}}{Z_{A}(\blam)^{-L}} \times \prod_{l=1}^{L} \frac{e^{-\beta H(\xb_{l}'(0);\blam')} }{e^{-\beta H(\xb_{l}(0);\blam)} }
 \\  \times \prod_{l=1}^{L} \frac{h_A(\xb_{l}'(0)) h_B(\xb_{l}'(\tobs))}{h_A(\xb_{l}(0)) h_B(\xb_{l}(\tobs))} \frac{P(\xvec_{l}'|\blam', \xb_{l}'(0))}{P(\xvec_{l}|\blam, \xb_{l}(0))}.
 \end{align}
As before in Section~\ref{sec:recZ}, we replace the reciprocal partition functions by unbiased estimates.
Specifically, \(Z_A(\blam)^{-L}\) is replaced by the product of \(L\) independent estimates of \(Z_A(\blam)^{-1}\), each computed as described in the previous sections:
\begin{equation}
Z_{A}(\blam)^{-L} \to \prod_{l=1}^{L} \widehat{Z_{A}(\blam)^{-1}}
\end{equation}
A demonstration that the unbiased estimates may be used in the acceptance probabilities follows in analogy to Eq.~\eqref{eq:lift}.
For each of the \(L\) estimates, one introduces a lift to include some noise variables \(\bdeta_l\).

\section{Results}
\label{sec:results}

To illustrate the design sampling with noisy estimates, we numerically studied the rate of escape from an energy well as a function of the well depth \(\epsilon\).
The potentially high-dimensional design \(\blam\) of Section~\ref{sec:theory} is just the scalar \(\epsilon\) for this application.
The toy problem was chosen to be sufficiently simple that brute force rate calculations \(\kAB(\epsilon)\) could also be collected to ensure that the procedure sampled designs---in this case well depths \(\epsilon\)---according to \(P(\epsilon) \propto \kAB(\epsilon)\).
Through numerical sampling, we confirmed that use of the noisy estimates \(\widehat{\tilde{Z}_A(\epsilon)^{-1}}\) do not bias the sampling.
We furthermore demonstrate that, however inconvenient to estimate, the partition function terms cannot be responsibly neglected; doing so yields a notable bias.

The specific toy model is the escape of a particle from from a Lennard-Jones well while evolving with underdamped Langevin dynamics in three dimensional space.
The energy well takes the familiar form
\begin{equation}
U(r; \epsilon) = 4 \epsilon \left[ \left(\frac{\sigma}{r}\right)^{12} - \left(\frac{\sigma}{r}\right)^6 \right],
\label{eq:LJ}
\end{equation}
where \(r\) is distance from the particle to the origin and \(\sigma\) is the particle radius.
We take the \(A\) region to be the bottom of the well, defined by positions with \(0.85 \leq r / r_{\rm min} \leq 1.4 \), with \(r_{\rm min} = 2^{1/6} \sigma\) being the location of the potential energy minimum.
The \(B\) region is the unbound state, reached once \(r\) exceeds \(4 \sigma\).
At every moment of time the particle experiences forces from this potential energy as well as a drag force and random fluctuating force from the underdamped Langevin dynamics.
Hence,
\begin{equation}
\dot{\mathbf{p}}(t) = -\nabla U(\mathbf{r}(t);\epsilon) - \frac{\gamma}{m} \mathbf{p}(t) + \bxi(t),
\label{eq:langevin}
\end{equation}
where \(\gamma\) is a friction coefficient and \(\bxi\) is a white noise with \(\left<\bxi\right> = \mathbf{0}\) and \(\left<\bxi(t) \bxi(t')\right> = 2 \gamma k_{\rm B} T \delta(t - t')\).
We allow our tunable design parameter \(\epsilon\) to vary between \(7\) and \(12 \ k_{\rm B} T\), a parameter regime chosen to ensure that escape is a rare event.
For convenience, we nondimensionalize the problem by setting \(\sigma = m = \beta = 1\) and \(\gamma = 0.5\).
This system is illustrated in Fig.~\ref{fig:Fig1}.

We split our results into four parts.
First we consider a joint Monte Carlo sampling of designs and particle positions to demonstrate that the noisy estimate for the reciprocal partition function can be adequately incorporated into a Monte Carlo acceptance ratio.
We next perform the joint sampling of designs and trajectories to extract the dependence of the rate constant on the well depth in the range \(\epsilon \in [7,12]\).
We demonstrate the enhanced preference for faster rate constants that comes from simultaneously sampling multiple reactive trajectories.
Finally, we illustrate the methodology on a more complex molecular cluster.

\subsection{Monte Carlo sampling with reciprocal partition function estimates}
\label{sec:phase}

As detailed in Section~\ref{sec:recZ}, the Monte Carlo sampling over trajectories and well depths involves the computation of an acceptance ratio, Eq.~\eqref{eq:rhoratio2}, containing \(\widehat{\tilde{Z}_A(\epsilon)^{-1}}\).
We note, however, that the reciprocal partition function enters this ratio not because of the trajectory sampling, but rather due to the sampling of the initial condition for the trajectory.
To evaluate the consequences of estimating \(\tilde{Z}_A(\epsilon)^{-1}\), we first chose to study a simpler subproblem: simultaneous sampling of well depths and initial positions (as opposed to full trajectories).

We constructed MCMC moves that transition from an old position and well depth, \(\mathbf{r}\) and \(\epsilon\), to a new position and depth, \(\mathbf{r}' = \mathbf{r} + \Delta \mathbf{r}\) and \(\epsilon' = \epsilon + \Delta \epsilon\).
The symmetric proposal is generated by drawing independent Gaussian variables \(\Delta r\) and \(\Delta \epsilon\), each with zero mean and variance \(10^{-4}\).
When a trial move generated \(\epsilon'\) outside the range \([7,12]\), the move was rejected.
Otherwise, the moves were accepted in one of three different ways:
according to the exact partition function
\begin{equation}
P^{\rm (exact)}_{\rm acc} = \min\left[1, h_{A}(\mathbf{r}')\frac{\tilde{Z}_{A}(\epsilon')^{-1} e^{-\beta U(\mathbf{r}';\epsilon')}} {\tilde{Z}_{A}(\epsilon)^{-1} e^{-\beta U(\mathbf{r};\epsilon)}}\right],
\label{eq:exact}
\end{equation}
according to the estimated partition function
\begin{equation}
P^{\rm (est)}_{\rm acc} = \min\left[1, h_{A}(\mathbf{r}')\frac{\widehat{\tilde{Z}_{A}(\epsilon')^{-1}} e^{-\beta U(\mathbf{r}';\epsilon')}} {\widehat{\tilde{Z}_{A}(\epsilon)^{-1}} e^{-\beta U(\mathbf{r};\epsilon)}}\right],
\label{eq:boothMC}
\end{equation}
or neglecting the partition function altogether
\begin{equation}
P^{\rm (ignored)}_{\rm acc} = \min\left[1, h_{A}(\mathbf{r}')\frac{e^{-\beta U(\mathbf{r}';\epsilon')}} {e^{-\beta U(\mathbf{r};\epsilon)}}\right].
\label{eq:ignored}
\end{equation}
We note that for Eq.~\eqref{eq:boothMC}, one must retain the old estimate for the reciprocal partition function after a rejection rather than recomputing a new estimate.
This need follows from the fact that we formally consider a random walk through the \(\eta\) coordinates that produced the noisy estimate, as described in Section~\ref{sec:recZ}

Equations.~\eqref{eq:exact} and~\eqref{eq:boothMC} should both sample marginal distributions for \(\epsilon\) that are uniform, while Eq.~\eqref{eq:ignored} samples \(P(\epsilon) \propto Z_A(\epsilon)\).
We confirmed these marginal distributions numerically by sampling particle positions in the Lennard-Jones well, a comparison made possible by the ease of numerically computing the exact partition function
\begin{equation}
\tilde{Z}_A(\epsilon) = 4 \pi \int_{0.85 r_{\rm min}}^{1.4 r_{\rm min}} dr \, r^2 e^{-\beta U(r; \epsilon)}
\end{equation}
for this toy model with \(U(r;\epsilon)\) given by Eq.~\eqref{eq:LJ}.

\begin{figure*}
  \centering
  \includegraphics[width=1.0\textwidth]{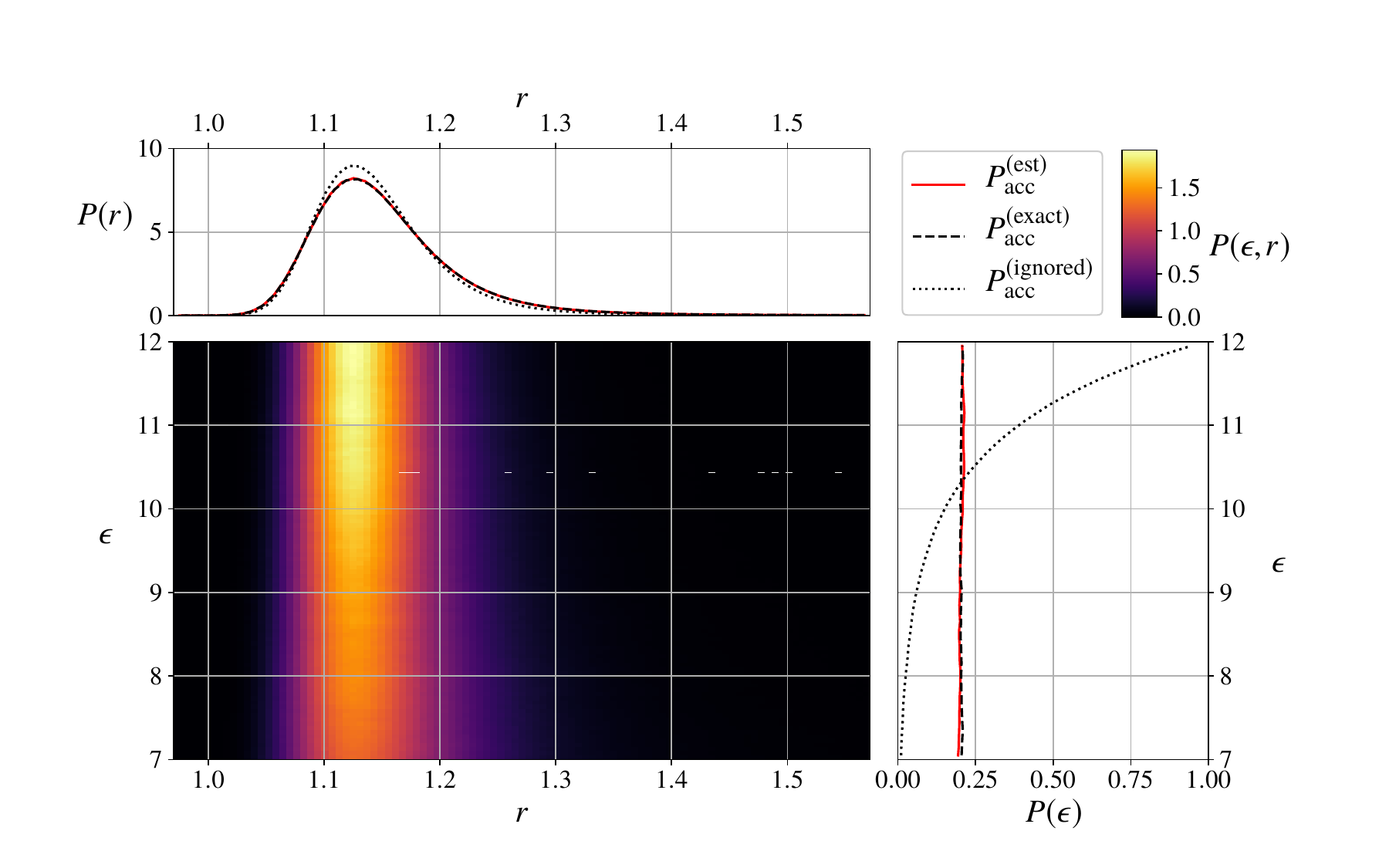}
  \caption{
    The joint probability densities of \(\epsilon\) and \(r\) for the sampling procedure described by Eq.~\eqref{eq:boothMC}.  
    Marginal probability densities for \(\epsilon\) and \( r\) are given along their respective axes from the joint distribution. 
    Marginal densities show data from the sampling procedure using reciprocal partition function estimates Eq.~\eqref{eq:boothMC} (red), the exact partition function Eq.~\eqref{eq:exact} (dashed black), and incorrectly ignoring the reciprocal partition function contributions Eq.~\eqref{eq:ignored} (dotted black). 
    Distributions were each collected from \(5 \times 10^8\) Monte Carlo trial moves, and the estimated reciprocal partition functions were computed with a roulette parameter of \(R=0.75\) and a reference value of \(\epsilon_{\rm ref}=12\).
    Marginal densities for the estimation procedure and the exact partition function agree well and the marginal density of \(\epsilon\) is uniform, as expected.
    Ignoring the reciprocal partition function introduces a non-negligible bias; rather than sampling \(\epsilon\) uniformly, it is sampled in proportion to to \(\tilde{Z}_{A}(\epsilon)\).
  }
  \label{fig:Fig2}
\end{figure*}

Fig.~\ref{fig:Fig2} gives the joint probability density from sampling in \(\epsilon\) and \( r\) space using reciprocal partition function estimation, Eq.~\eqref{eq:boothMC}. 
The marginal distributions in \(\epsilon\) and \(r\) are shown along their respective axes and give comparisons with sampling using the exact partition function, Eq.~\eqref{eq:exact}, and ignoring the partition function contributions, Eq.~\eqref{eq:ignored}.
We can see that using unbiased reciprocal partition function estimation works well, matching the results obtained from the exact partition functions.
Both the estimated and exact approaches also result in a uniform marginal distribution across \(\epsilon\), as expected.
By carefully constructing an unbiased estimation procedure, we have recovered the proper sampling without having to laboriously calculate an exact partition function at every sampled value of \(\epsilon\).
The marginal distribution of \(\epsilon\) also shows the effect of the reciprocal partition function ratio on the sampling procedure.
By ignoring this ratio, we sample \(\epsilon\) in proportion to \(\tilde{Z}_{A}(\epsilon)\), which introduces a bias that prefers higher values of \(\epsilon\).
The non-uniform distribution reflects the fact that trajectory sampling would show a preference for some designs not only because of the propensity to react, but also due to the ease of generating initial configurations.
From Fig.~\ref{fig:Fig2}, it is clear that this bias can be significant.

Of course to sample the joint distribution in Fig.~\ref{fig:Fig2}, we need to generate estimates for the reciprocal partition functions.
The first step in doing so is to choose an appropriate reference \(\epsilon_{\rm ref}\) following the considerations of Section~\ref{sec:refdist}.
For the range \(\epsilon \in [7,12]\) a value of \(\epsilon_{\rm ref} = 12\) gives the lowest energy at any value of \(r\) due to the monotonically decreasing energy of the Lennard-Jones form with increasing \(\epsilon\).
There is no need for an offset \(U_{0}\).
Using standard MCMC for fixed \(\epsilon_{\rm ref}\) we generated 10,000 independent samples of position \(\mathbf{r}\) and corresponding reference energy \(U(\mathbf{r};\epsilon_{\rm ref})\) to construct a library.
As described in Section~\ref{sec:refdist}, we generated \(a^{(i)}\) terms in the expansion of Eq.~\eqref{eq:truncseries} for arbitrary \(\epsilon\) by uniformly randomly drawing samples from the library and evaluating Eq.~\eqref{eq:a}.
The estimates of Fig.~\ref{fig:Fig2} were generated with a roulette parameter of \(R = 0.75\).

\begin{figure*}[ht]
  \centering
  \resizebox{1.0\textwidth}{!}{%
  \includegraphics[height=6cm]{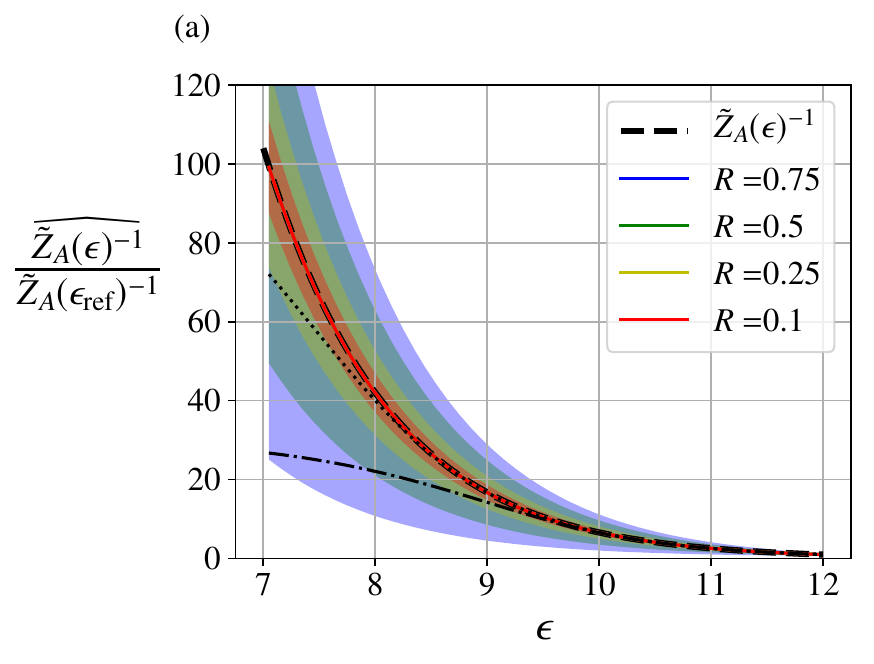}%
  \quad
  \includegraphics[height=6cm]{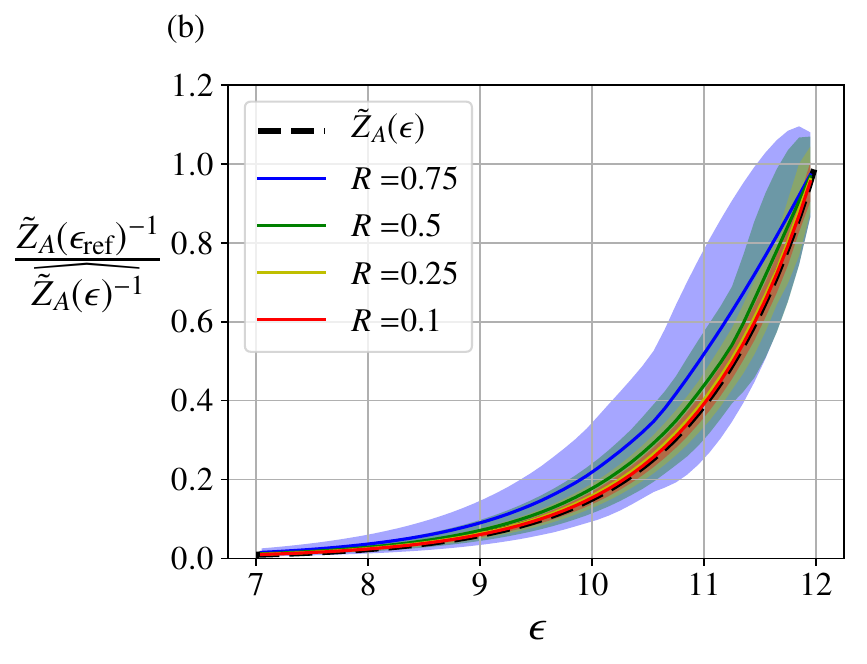}%
  \quad \quad \quad
  \includegraphics[height=6cm]{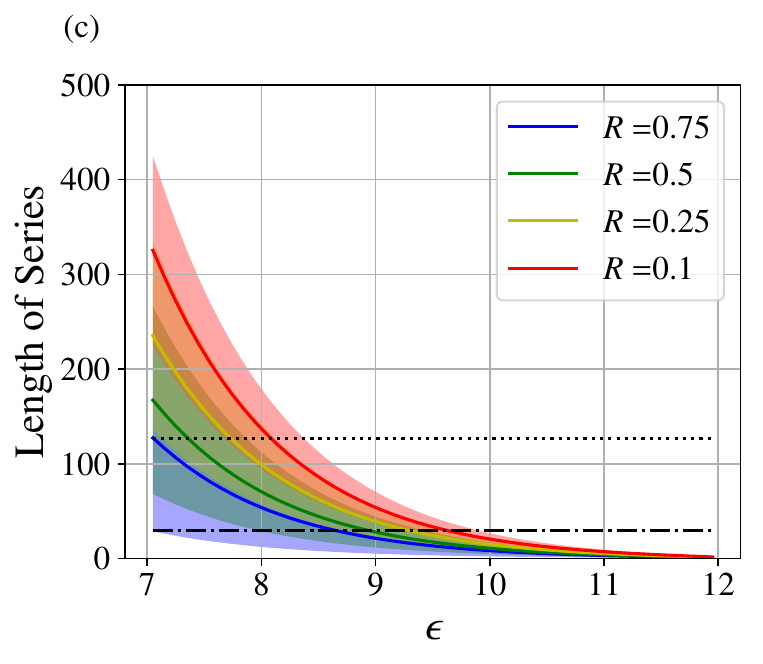}%
  }
  \caption{
  (a) Unbiased estimates of the reciprocal partition function \( \widehat{ \tilde{Z}_{A}(\epsilon)^{-1} } \).
  (b) Biased estimates of the partition function \(\tilde{Z}_{A}(\epsilon)\) constructed by taking the reciprocal of \( \widehat{ \tilde{Z}_{A}(\epsilon)^{-1} } \).
  Both estimates are scaled by the value of the exact partition function at the reference \(\epsilon_{\rm ref} =12\) and the exact partition function is given as a thick dashed black line.
  (c) The series length when using the Booth method as a function of the well depth \(\epsilon\).  
  Different colors represent different roulette parameters with the solid lines giving the average value and the shaded areas showing the standard deviation.
To emphasize the importance of stochastic truncation, biased results are shown for series truncated at fixed lengths of 30 and 127 terms, plotted as dash-dotted and dotted lines, respectively.
  Average estimates of the reciprocal partition function are indistinguishable from the exact result in (a), demonstrating the unbiased nature of the Booth estimation procedure.
  The average estimate values only approach the exact values in the limit \(R \to 0\) in (b), showing that the reciprocal of our unbiased estimates for \(\tilde{Z}_{A}^{-1}\) does not yield an unbiased estimate of \(\tilde{Z}_{A}\).
  From (c) we can see that the cost of the estimation method increases as \(R\) decreases and also as the value of \(\epsilon\) strays further from the reference \(\epsilon_{\rm ref} = 12\).
  }
  \label{fig:Fig3}
\end{figure*}

Figure~\ref{fig:Fig3}(a) shows that the reciprocal partition function remains unbiased regardless of the choice of \(R\), but that \(R\) impacts the estimator's noise.
The figure also highlights that the stochastic truncation of the series expansion is essential and that termination at a fixed expansion length would yield biased estimates.
Likewise, the reciprocal of our stochastic series is a \emph{biased} estimate of the partition function itself, as shown in Fig.~\ref{fig:Fig3}(b).
That bias is expected and not problematic; our MCMC scheme required unbiased \(\tilde{Z}_{A}^{-1}\), not unbiased \(\tilde{Z}_{A}\).
While the procedure formally works irrespective of \(R\) or the distance from the reference potential, the cost of the estimation procedure varies.
Fig.~\ref{fig:Fig3}(c) shows that decreasing \(R\) decreases the probability of truncation, resulting in a stochastic sum with more terms.
A series with more terms is better converged since it effectively averages over many more values of \(a^{(i)}\), but the decrease in the noise comes at a computational expense.

Tuning \(R\) to select an optimal trade-off between noise and computational cost is a complicated affair.
One advantage of using estimates to sample is that one can get by with noise, potentially very large noise without introducing bias, suggesting that one should favor very cheap, noisy estimates.
However, very noisy estimates can cause the Markov chain to get stuck in \(\bdeta\) variables that produce an overly favorable estimate.
Practical implementations require care---in choosing \(R\), in selecting a reference potential, and in preventing stuck Markov chains---but our calculations serve as a demonstration of the principle that the noisy estimates of reciprocal partition functions can be computed and productively employed.

\subsection{Simultaneously sampling trajectory space and design space}
\label{sec:traj}
Having demonstrated the ability to sample the design and the initial condition, we now want to bias designs so as to favor fast rates.
Section~\ref{sec:theory} laid out two routes to sample \(P(\epsilon) \propto \kAB(\epsilon)\).
If we could compute \(\tilde{Z}_A^{-1}\) exactly, we could sample designs and trajectories according to Eq.~\eqref{eq:rho} with a MCMC procedure that updates a trajectory \(\xvec\) and a design \(\epsilon\).
Otherwise, we could also generate estimates for the reciprocal of the partition function to sample Eq~\eqref{eq:lift}.
As in Section~\ref{sec:phase}, we consider the Lennard-Jones escape problem because it is reasonable to implement both routes---using exact and estimated reciprocal partition functions---as a demonstration of validity.

Both routes require a random generation of new designs and trajectories, \(P_{\rm gen}(\xvec, \epsilon \to \xvec', \epsilon')\).
We make these proposals in two steps.
First, we symmetrically generate \(\epsilon' = \epsilon + \Delta \epsilon\) as in Section~\eqref{sec:phase}.
Next, we use \(\epsilon'\) to generate a new trajectory via a ``shooting move'' that re-evolves the stochastic dynamics forward and backward in time from a randomly selected time~\cite{bolhuis2002transition}.
The combined move is conditionally accepted according to Eq.~\eqref{eq:pacc} (exact) or to Eq.~\eqref{eq:pacc2} (est), both of which require an explicit calculation of the ratio of generation probabilities, as well as a measure of the Boltzmann probability of the new trajectory's initial condition.
That Boltzmann probability of the initial condition is handled as in Section~\ref{sec:phase}---we either compute it exactly (exact) or we generate an unbiased estimate (est).
Unlike Section~\ref{sec:phase}, we now must compute the ratio of the trajectory generation probabilities, a ratio that can be computed explicitly in terms of the random noise terms for the stochastic dynamics.
Appendix~\ref{sec:trajapp} provides details of the trajectory generation probability based on the underdamped Langevin integrator of Ath\`enes and Adjanor~\cite{athenes2008measurement}.

\begin{figure}[ht]
  \centering
  \includegraphics[width=0.55\textwidth]{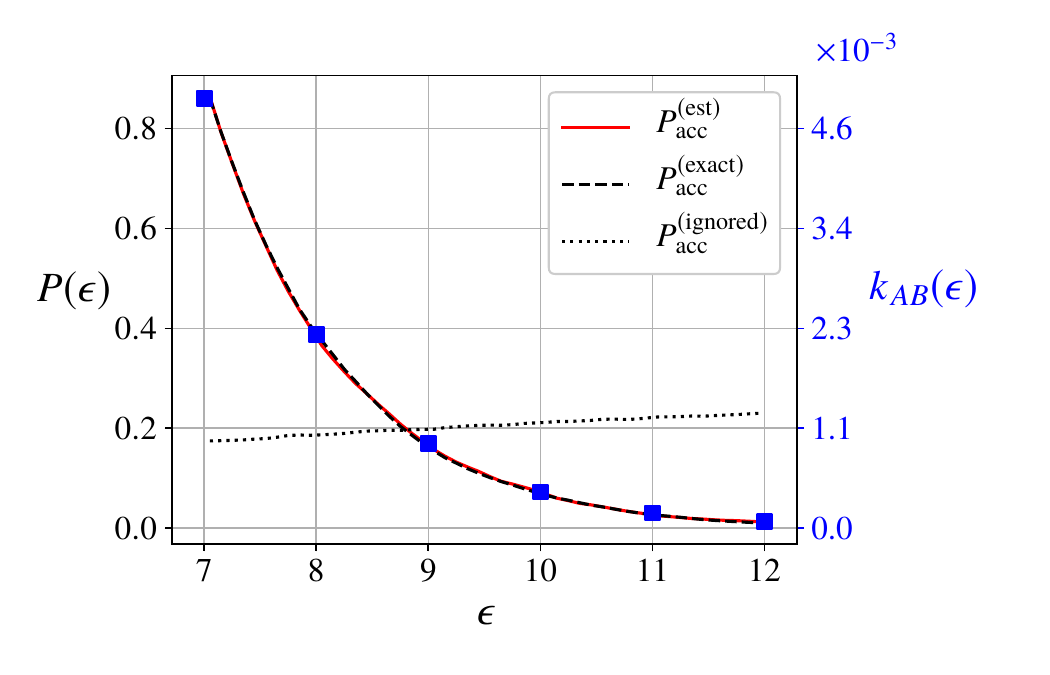}
  \caption{ (left, black axis) Marginal probability densities of \( \epsilon \) from simultaneous sampling of the design space \(\epsilon\) and of unbinding trajectories.
  Sampling using the exact reciprocal partition functions Eq.~\eqref{eq:pacc} gives the dashed black curve and sampling with estimated reciprocal partition functions Eq.~\eqref{eq:pacc2} gives the solid red curve.
  (right, blue axis) Brute force estimations of the average rate constants of unbinding \( \kAB(\epsilon) \) as a function of \( \epsilon\) given as blue squares. 
  Each trajectory sampling procedure used \(5 \times10^8\) trial attempts and the estimation procedure used a roulette parameter \(R=0.75\) and a reference value of \(\epsilon_{\rm ref}=12\).
  Each rate calculation consisted of \(10^6\) independent trials.
  To overlay \(P(\epsilon)\) and \( \kAB(\epsilon) \), the proportionality constant was fitted by least squares regression.
  }
  \label{fig:Fig4}
\end{figure}

To confirm that the trajectory sampling approach yields the rate of unbinding, we additionally computed the rate constant as a function of well depth \(\kAB(\epsilon)\) by brute force.
For those brute force calculations, we used MCMC to initialize the Lennard-Jones particle's position and momentum from the equilibrium Boltzmann distribution then propagated the trajectory~\cite{athenes2008measurement} with a time step of \(\Delta t = 0.005\).
Once the system reached \(B\) we recorded the elapsed time \(\tau\) and repeated the procedure.
In all we sampled \(10^{6}\) realizations of \(\tau\) and calculated an estimate of \( \kAB \) as \(1/\langle \tau \rangle\).
In Fig.~\ref{fig:Fig4} we overlaid the rate constant data on top of the marginal distributions of \(\epsilon\) taken from trajectory sampling.

Fig.~\ref{fig:Fig4} shows that using our reciprocal partition function estimation procedure works well, as it matches the data collected when using an exact partition function.
Furthermore, by overlaying the rate constant data we show that we are indeed sampling \(\epsilon\) in proportion to the rate constant, \(P(\epsilon) \propto \kAB(\epsilon)\).
Consequently, the random walkers executing both the exact and estimation approaches spend most of their time sampling designs with fast rates \(\epsilon \approx 7\).
In contrast, the random walker that ignores the partition function factor in its acceptance ratio spends most of its time sampling the slow designs, those with \(\epsilon \approx 12\).
These data confirm that the partition function term can be very significant in rate design problems, and that it can be computed approximately in manner that avoids sampling bias.

\subsection{Sampling with stronger bias for fast rates}
\label{sec:multi}
As discussed in Section~\ref{sec:multitraj}, visiting designs in proportion to their rates is a relatively weak preference in favor of fast rates.
In high dimensional design spaces, the design entropy overwhelms that weak preference, so we sought a way to turn up the bias by sampling \(P(\epsilon) \propto k(\epsilon)^L\) with \(L\) independent reactive trajectories.
Implementing this scheme follows quite directly from the previous section.
The principal difference is that a MC move in \(\epsilon\) and \(\xvec\) now becomes a move in \(\epsilon, \xvec_1, \xvec_2, \hdots \xvec_L\).
Specific computational details are discussed in Appendix~\ref{sec:trajapp}.

Figure~\ref{fig:Fig5} summarizes the result of sampling with \(L = 1, 2,\) and \(3\).
The fastest rate of escape occurs with the shallowest allowed well, \(\epsilon = 7\), but with \(L = 1\) there is still appreciable probability of seeing \(\epsilon\) fluctuate to suboptimal values above 9 or 10.
By increasing \(L\), the density near the optimal \(\epsilon\) grows.
Figure~\ref{fig:Fig5}(a) shows that sampling using the estimated reciprocal partition functions remains unbiased for \(L > 1\).
The more stringent confirmation that \(P(\epsilon) \propto k(\epsilon)^L\), or equivalently that \(\log P(\epsilon)\) versus \(\log k(\epsilon)\) has slope \(L\), is plotted in Fig.~\ref{fig:Fig5}(b-d).

\begin{figure}[ht]
  \centering
  \includegraphics[width=0.45\textwidth]{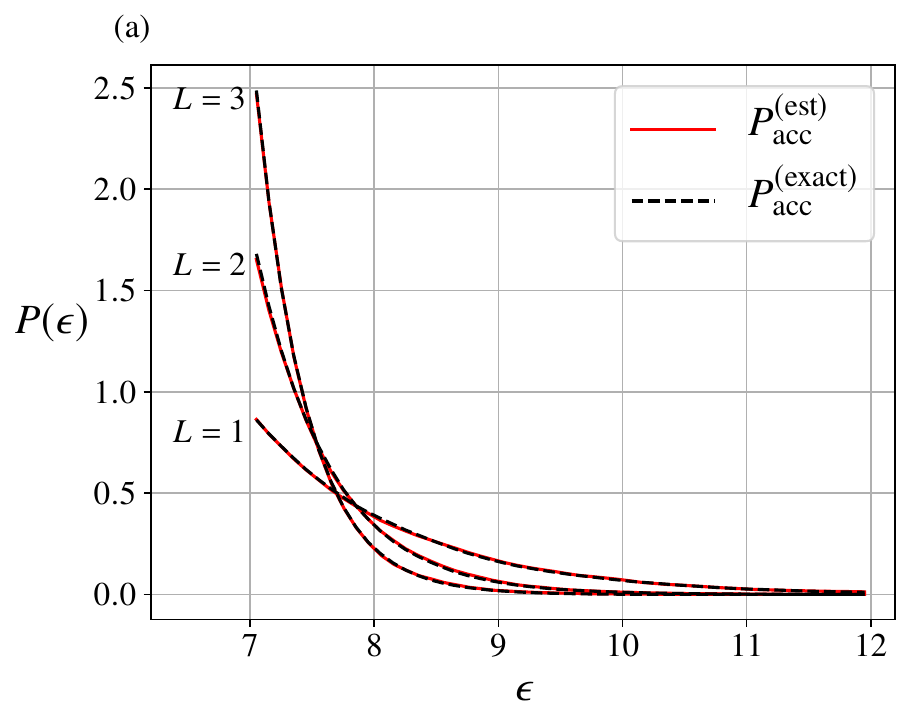}
  \includegraphics[width=0.5\textwidth]{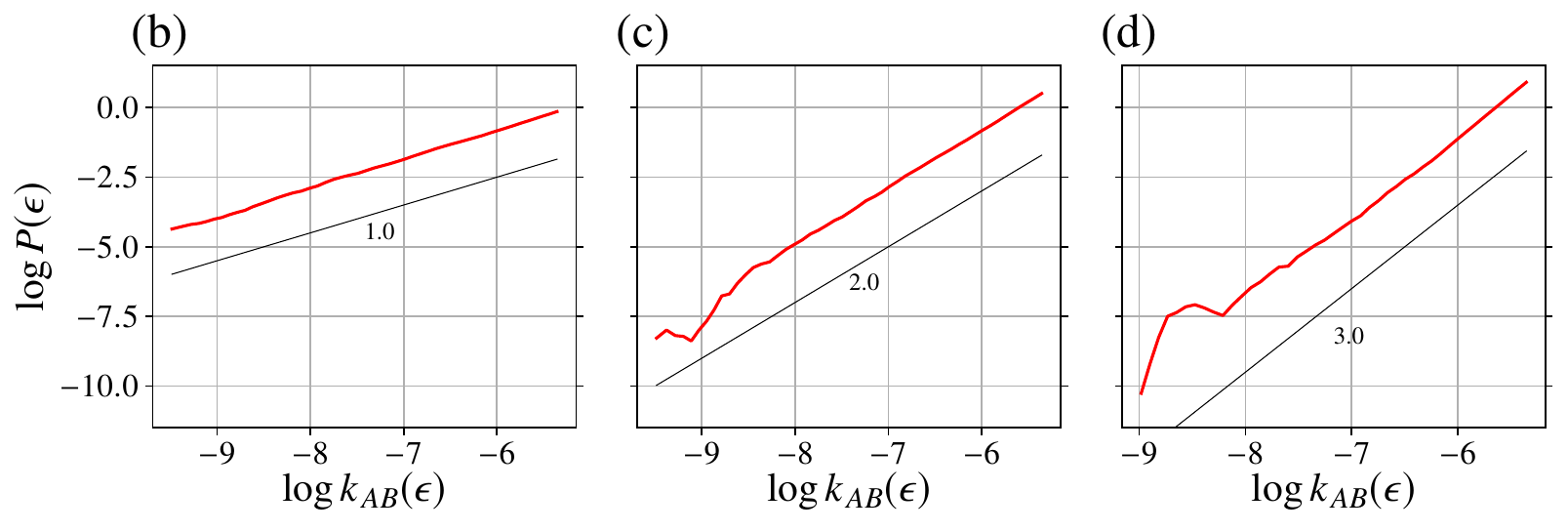}
  \caption{(a) Marginal probability densities of \(\epsilon\) from joint sampling of the design space \(\epsilon\) and of \(L\) independent unbinding trajectories, as described in Section~\ref{sec:multitraj}.
  Accounting for the reciprocal partition function exactly is shown in dashed black and using the reciprocal partition function estimations in solid red. 
  (b) The marginal probability densities from the estimation procedure in (a) are plotted against the brute force calculated rate constants (blue squares in Fig.~\ref{fig:Fig4}) on a log-log plot for (b) \(L=1\), (c) \(L=2\), and (d) \(L=3\).  Power law scalings are provided in solid black as reference. 
  The brute force rate constants were fitted to a high-order polynomial for interpolation purposes.
  Each trajectory sampling procedure used \(5 \times 10^8\) trial attempts, and the estimates used a roulette parameter of \(R=0.75\) and a reference value of \(\epsilon_{\rm ref} =12\).
  From (a) we can see that increasing the number of independent trajectories in the sampling creates a stronger preference for the fastest unbinding rate, which occurs at \(\epsilon = 7\).
  From (b), (c), and (d) we conclude that we are sampling according to \(P(\epsilon) \propto \kAB(\epsilon)^{L}\).
  The power law scaling at low values of \(\kAB\) (high values of \(\epsilon\)) is noisy for larger \(L\) values because slow values of \(\kAB\) are rendered particularly rare by using a large number of independent trajectories.
  }
  \label{fig:Fig5}
\end{figure}

\subsection{A more complex example}
\label{sec:complex}

\begin{figure}[ht]
  \centering
  \includegraphics[width=0.43\textwidth]{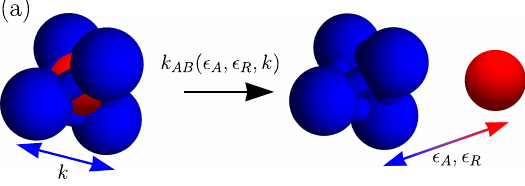} 
  \includegraphics[width=0.5\textwidth]{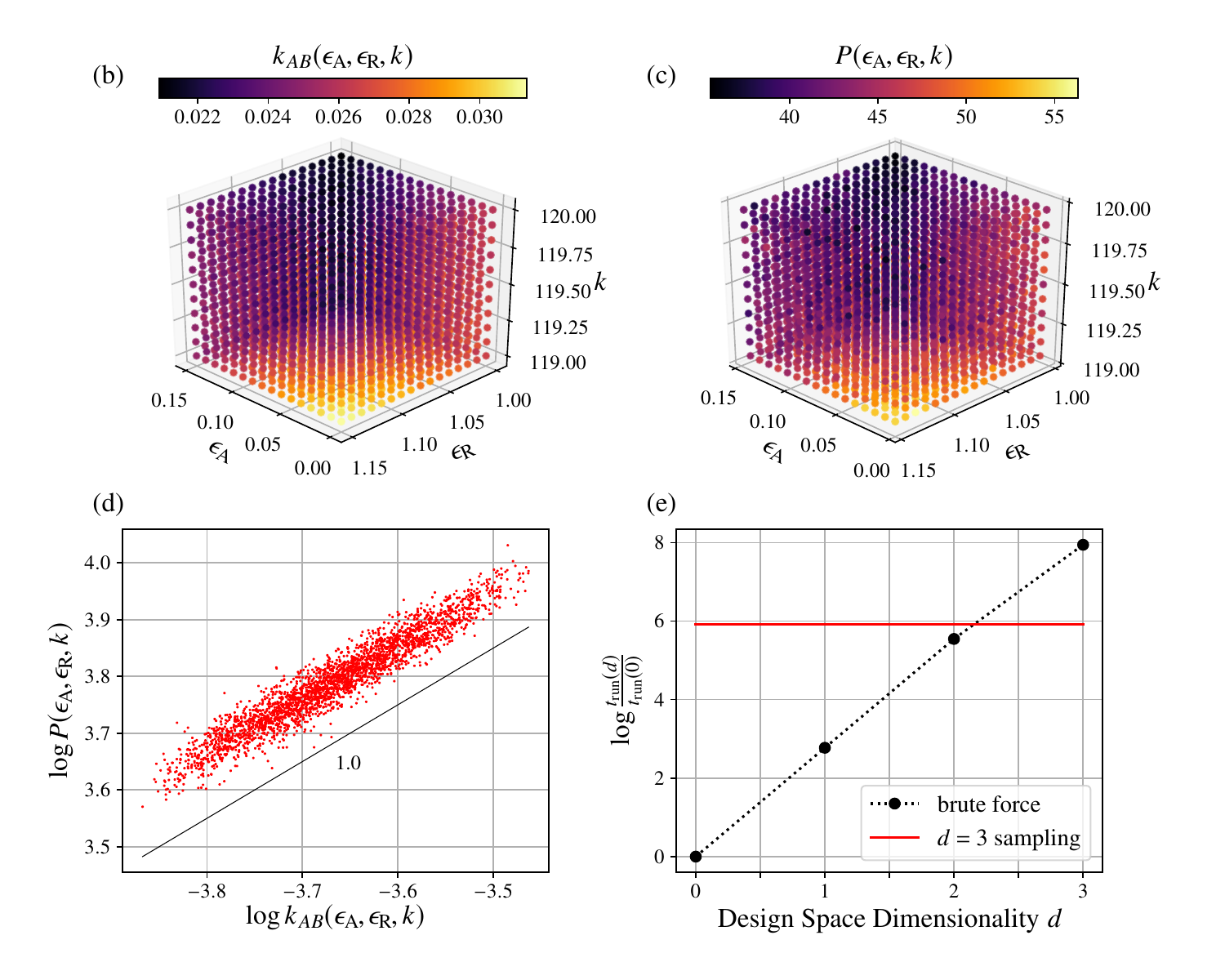}
  \caption{ (a) A tetrahedral cluster consisting of five Lennard-Jones particles releases the trapped red particle from its blue cage with a rate that depends on three parameters of the pairwise interactions.
   (b) Brute force rate calculations for a grid of designs, estimated as in the Lennard-Jones unbinding example as the reciprocal of the average escape time of \(2 \times 10^4\) independent underdamped Langevin trajectories.
   (c) Binned probability density for design sampling, revealing relative rate constants that match the brute force calculation.
Design sampling utilized trajectories with \(M=1000\) time steps, reciprocal partition function parameter \(R = 0.75\), and reference design \(\epsilon_{\rm A} = 0.15\), \(\epsilon_{\rm R} = 1\), and \(k = 119\) that was sampled by \(10^6\) reference configurations.
Trajectories were sampled as described in Appendix~\ref{sec:trajapp}.
    (d) A quantitative comparison of the brute force rates and the sampled designs.
As in Fig.~\ref{fig:Fig5}(b), the unit slope on the log-log scale reflects equality up to a proportionality constant.
     (e) Sampled designs can be generated with a cost independent of design dimensionality \(d\) while brute force rate computations scale exponentially with \(d\).
Though we never aimed to fully optimize the sampling strategy, the runtime \(t_{\rm run}\) to generate (d) was already below that of (c) for the three-dimensional design problem.}
  \label{fig:Fig6}
\end{figure}

While the Lennard-Jones unbinding system is useful for demonstrating correctness, the power of the sampling approach is most applicable to more complex problems with high-dimensional state spaces and design spaces. 
To illustrate such an application, we introduce a five particle system shown in Fig.~\ref{fig:Fig6} that consists of one particle metastably trapped inside a tetrahedron of four other particles.
Given enough time, the trapped particle escapes from the tetrahedron with a rate constant that depends on the various interactions between particles.
Whereas the Lennard-Jones unbinding example supported a single trivial reaction pathway, competing reaction pathways are possible with multiple particles.
In this respect, this toy model serves as a more realistic example that builds toward chemical applications.

In addition to increasing the dimensionality of the state space, we now consider a three-dimensional design space which regulates interparticle interactions.
One of those design degrees of freedom adjusts a spring constant \(k\) for a harmonic potential acting between pairs of the tetrahedron particles.
When separated by a distance \(r_{ij}\) this pair potential contributes \(U_{\mathrm{harmonic}}(r_{ij} ; k) = \frac{1}{2} k r_{ij}^{2}\) to the energy.  
All five particles additionally interact through modified Lennard-Jones potentials,
\begin{equation}
U_{\rm LJ}(r_{ij} ; \epsilon_{\rm R}, \epsilon_{\rm A}) = 4 \epsilon_{R} \left( \frac{\sigma_{ij}}{r_{ij}} \right)^{12} - 4 \epsilon_{A} \left( \frac{\sigma_{ij}}{r_{ij}} \right)^{6}  .
\label{eq:modLJ}
\end{equation}
Here \(\sigma_{ij}\) is the mean of the Lennard-Jones radii of particles \(i\) and \(j\).
We call the potentials modified Lennard-Jones potentials because we allow attractive and repulsive interactions to be respectively tuned via \(\epsilon_{\rm A}\) and \(\epsilon_{\rm R}\).
In principle those \(\epsilon_{\rm A}\) and \(\epsilon_{\rm R}\) parameters could be distinct for each pairwise interaction, but we wanted to limit to a three-dimensional design space in order to visually compare with brute force calculations.
Thus, we held fixed \(\epsilon_{\rm A} = 0\) and \(\epsilon_{\rm R} = 10\) between tetrahedron particles and took \(\epsilon_{\rm A}\) and \(\epsilon_{\rm R}\) between the trapped particle and each tetrahedron particle to be the two remaining design parameters.
All other parameters were held fixed: \(k_{\rm B} T = 1\), and \(\gamma = 0.5\), \(\Delta t = 0.005\), unit mass for all particles, unit radii for all tetrahedron particles, and a trapped particle radius of 0.45.
The design \(\blam = \left\{ k, \epsilon_{\rm A}, \epsilon_{\rm R}\right\}\) was allowed to sample, restricted to the domain with \(\epsilon_{\rm R} \in [1, 1.15]\), \(\epsilon_{\rm A} \in [0, 0.15]\), and \(k \in [119, 120]\).
Trial moves in parameter space were generated via \(\epsilon_{\rm A}^\prime = \epsilon_{\rm A} + \Delta \epsilon_{\rm A}\), \(\epsilon_{\rm R}^\prime = \epsilon_{\rm R} + \Delta \epsilon_{\rm R}\), and \( k^\prime = k + \Delta k \), where \( \Delta \epsilon_{\rm A}\), \(\Delta \epsilon_{\rm R}\), and \(\Delta k\) where drawn from Gaussian distributions with 0 mean and variances of \(10^{-4}\), \(10^{-4}\), and \(10^{-2}\), respectively.
The criteria for detecting reactant and product states was based on the distance between the center of mass of the tetrahedron and the trapped particle.
When that distance is below 0.25, the configuration is in the reactant state \(A\), and when it exceeds 1.1, the configuration is in the product state \(B\).

Figure~\ref{fig:Fig6} demonstrates the merits of the Monte Carlo methodology for this three-dimensional design problem.
The brute force rate calculations of Fig.~\ref{fig:Fig6}(b) reveal relative rates between designs which are essentially indistinguishable from the design sampling of Fig.~\ref{fig:Fig6}(c).
In both cases, the rate of the trapped particle's escape is maximized when the tetrahedron is the least rigid (low \(k\)) and when the interactions between the tetrahedron particles and the trapped particle are most repulsive and least attractive (high \(\epsilon_{\rm R}\), low \(\epsilon_{\rm A}\)).  
The proportionality between brute force rate and design sampling is made quantitative by comparing \(\log P(\blam)\) with \(\log k_{AB}(\blam)\) in Fig.~\ref{fig:Fig6}(d).

The computational benefit of the sampling procedure becomes apparent in this higher dimensional parameter space.
Fig.~\ref{fig:Fig6}(e) makes this benefit clear by comparing run times for computing rates in \(d\) dimensions with \(d=0\) corresponding to a single design, \(d=1\) to \(\blam = \left\{\epsilon_{\rm A} \right\} \), \(d=2\) to \(\blam = \left\{\epsilon_{\rm A}, \epsilon_{\rm R} \right\} \), and \(d=3\) to \(\blam = \left\{\epsilon_{\rm A}, \epsilon_{\rm R}, k \right\} \).
The cost for performing calculation on a grid like this scales exponentially, whereas the sampling procedure's cost depends only on the number of sampled designs, not the design dimensionality.
Because of the poor scaling of grids, the sampling procedure eventually wins out as dimensionality increases, a crossover that occurred after \(d=2\) for our implementation.

\section{Discussion}
\label{sec:discussion}
In this manuscript, we have demonstrated how to sample design spaces with a preference for fast reaction rates.
Our central focus has been the development of a practical Monte Carlo strategy that samples \(p(\blam) \propto \kAB(\blam)^L\) without bias.
Our success studying toy models demands a level-headed assessment of whether the methodology will scale to more complex problems with still higher dimensional \(\blam\).
As a Monte Carlo strategy, there is reason to believe that high-dimensional spaces could be accessible, but here we must highlight two reasons for caution.

One concern is that our unbiased estimates of \(Z_A(\blam)^{-1}\) came from a comparison against a single fixed reference with energy \(U(\mathbf{r}; \blam_{\text{ref}}) + U_0\).
Like the estimation of free energy differences from importance sampled configurations, efficient computations rely on good overlap with the reference distribution.
In our case, we require the typical \(\mathbf{r}\) sampled by \(U(\mathbf{r}; \blam)\) to be similar to those typically sampled by the reference potential.
Adequate overlap was simple to achieve in the toy problems due to our focus on \(d \le 3\) design dimensions.
We expect greater difficulty in higher dimensions, where it may be necessary to generate unbiased estimates of \(Z_A(\blam)^{-1}\) using samples from multiple different reference potentials.
We expect the multistate Bennett Acceptance Ratio method for unbiased free energy calculations likely guides the way toward simultaneously weighting multiple references~\cite{shirts2008statistically}.

Another concern one might raise is that the sampling procedure directly provides an ensemble of decent designs rather than a single optimal design, yet often it is this optimal design that is desired.
The situation is analogous with a finite-temperature canonical ensemble returning configurations that differ from the energy minima.
To discover those minima, it is necessary to quench by progressively lowering the temperature.
For the design sampling problem we showed that quench could be achieved by increasing \(L\), the number of random walkers.
However, the principal virtue of the method is that it allows one to importance sample design space, not just perform quenches.
That importance sampling promises insight going beyond the identification of an optimal design.
For example, it should be possible to assess how many near-optimal designs exist and by how much their performance erodes relative to the optimum.

\section{Acknowledgments}
This work used the Extreme Science and Engineering Discovery Environment (XSEDE) Comet cluster at the San Diego Supercomputer Center through startup allocation TG-CHE190024. TRG gratefully acknowledges a long history of useful conversations and guidance from Michael Gr\"{u}nwald and Phillip Geissler. TRG and AA thank Geyao Gu for many productive discussions.

\section{Data Availability}
The data and code that support the findings of this study are openly available at http://doi.org/10.5281/zenodo.4150012~\cite{code}.

\appendix
 
\section{Generation of new trajectories}
\label{sec:trajapp}

The Monte Carlo procedure of Sec~\ref{sec:recZ} expresses the probability of accepting a MC move in terms of a ratio of generation probabilities \(P_{\rm gen}(\xvec, \blam \to \xvec', \blam') / P_{\rm gen}(\xvec', \blam' \to \xvec, \blam)\).
Here we discuss details of the transition path sampling moves that allow us to compute the acceptance ratio and efficiently sample the trajectory space.
We focus on sampling stochastic dynamics, specifically discretized underdamped Langevin dynamics, in which case shooting moves combined with the strategy of Crooks is effective~\cite{crooks2001efficient}.
The trajectory \(\xvec\) consists of \(M+1\) snapshots of phase space, \(\mathbf{x}=\{\mathbf{r},\mathbf{p}\}\), separated by time step \(\Delta t \).
It is convenient to have a compact notation, so the \(j^{\rm th}\) snapshot occurring at time \(j \Delta t\), \(\mathbf{x}(j\Delta t)\), will simply be written as \(\mathbf{x}^j\).
The trajectory \(\xvec\) is defined either by this set \(\{\mathbf{x}^j\}\) for all \(j = 0, \hdots, M\) or alternatively by the initial condition \(\mathbf{x}^0\) and the set of all random numbers that cause the stochastic integrator to visit the subsequent points in phase space.

In Section~\ref{sec:traj} we employed the Langevin integrator of Ath\`enes and Adjanor~\cite{athenes2008measurement}, in which case each degree of freedom requires two Gaussian random variables (noises) to be drawn for each time step.
Similar to the velocity Verlet algorithm, the position coordinates defined at integer time steps are computed using velocities that are defined at integer and half-integer times.
The two noises for each degree of freedom can also be associated with fractional time steps, so the update of particle \(i\) can be computed as
\begin{align}
\nonumber \mathbf{p}_{i}^{j + \frac{1}{2}} &= \mathbf{p}_{i}^{j} e^{- \frac{\gamma \Delta t}{2 m_{i}}}  + \mathbf{f}_{i}^{j} \frac{\Delta t}{2} + \bxi_{i}^{j + \frac{1}{2}}\\
\nonumber \mathbf{r}_{i}^{ j+1}  &= \mathbf{r}_{i}^{j} + \mathbf{p}_{i}^{j + \frac{1}{2}} \frac{\Delta t}{m_{i}}\\
\mathbf{p}_{i}^{j + 1} &= \left[ \mathbf{p}_{i}^{j + \frac{1}{2}}  + \mathbf{f}_{i}^{j + 1} \frac{\Delta t}{2} \right] e^{-\frac{\gamma \Delta t}{2 m_{i} }} + \bxi_{i}^{j + 1},
\label{eq:athenes}
\end{align}
where the Gaussian white noises \(\bxi_{i}^{j + \frac{1}{2}}\) and \(\bxi_{i}^{j + 1}\) both have mean zero and variance \(m_{i}(1 - e^{- \frac{\gamma \Delta t}{m_{i} }})/\beta\).
The trajectory can then be expressed as \(\xvec \equiv \{ \mathbf{x}(0),\bxi^{\frac{1}{2}}, \bxi^{1}, \ldots, \bxi^{j + \frac{1}{2}}, \bxi^{j+1}, \ldots, \bxi^{M-\frac{1}{2}}, \bxi^{M} \} \).
We can also define the time-reversed trajectory as \(\xvecback \equiv \{\mathbf{x}(M\Delta t),\bxit^{M},\bxit^{M-\frac{1}{2}}, \ldots, \bxit^{j+1}, \bxit^{j + \frac{1}{2}}, \ldots, \bxit^{1}, \bxit^{\frac{1}{2}} \} \), where \(\bxit\) are the random numbers that would give a reversed trajectory through phase space starting from the endpoint with reversed momenta. 

The MC proposal \(\xvec, \blam \to \xvec', \blam'\) is thus constructed as follows.
First, we generate \(\blam \to \blam'\) symmetrically, and if a \(\blam'\) value is chosen outside the desired domain then the entire move is rejected.
Next, we randomly choose a ``shooting point'' \(m \in [0,M]\) along the trajectory with uniform probability.
From this point we modify the time-reversed random numbers from that point backwards to the beginning of the trajectory, \( \bxit^{j + \frac{1}{2}} \to \bxit^{\prime j + \frac{1}{2}} \) and \( \bxit^{j + 1} \to \bxit^{\prime j + 1} \) for \(j = 0,1,\ldots, m-1\).
Similarly we modify the forward time random numbers from that point to the end of the trajectory,  \( \bxi^{j + \frac{1}{2}} \to \bxi^{\prime j + \frac{1}{2}} \) and \( \bxi^{j + 1} \to \bxi^{\prime j + 1} \) for \(j = m, m+1, \ldots, M-1\).
We generate trial move noises by a linear combination of the old noise and a new one \( \bxi^\prime =  \alpha \bxi +\sqrt{1-\alpha^2}\bzeta \) and \( \bxit^\prime =  \alpha \bxit +\sqrt{1-\alpha^2}\bzetat \)~\cite{stoltz2007path,gingrich2015preserving}. 
Here  \(\bzeta\) and \(\bzetat\) are the new noises, drawn from the same zero mean, \(m_{i} (1 - e^{- \frac{ \gamma \Delta t}{m_{i}}})/\beta\) variance Gaussian distribution that the Langevin integrator noises naturally sample.
The parameter \(\alpha\) allows us to control the decorrelation between the current and trial trajectory. 
Starting from the random point \(m\) we integrate the trajectories backward in time using the proposed backward random numbers \(\bxit'\) and forward in time using the proposed forward random numbers \(\bxi'\) to generate the trial trajectory \(\xvec'\).
The resulting trajectory can then be converted to its equivalent complete forward and reverse time representations as random numbers, \(\bxi'\) and \(\bxit'\), respectively, using the form of the integrator in Eq.~\eqref{eq:athenes}.  

The reversed proposal move, \(\xvec', \blam' \to \xvec, \blam\), occurs when \(\blam' \to \blam\) is generated, the same shooting point is chosen, and \(\bzeta' = (\bxi - \alpha \bxi')/ \sqrt{1-\alpha^2}\) and \(\bzetat' = (\bxit - \alpha \bxit') / \sqrt{1-\alpha^2}\) are chosen from a Gaussian distribution to map from \(\bxi'\) back to \(\bxi\).
The relative probability of the forward move to the reverse is thus
\begin{widetext}
\begin{align}
\nonumber \frac{P_{\rm gen}(\xvec, \blam \to \xvec', \blam')}{P_{\rm gen}(\xvec', \blam' \to \xvec, \blam)} &= \exp\left[\sum_{i=1}^N \frac{\beta}{2 m_{i} (1 - e^{- \frac{ \gamma \Delta t}{m_{i}}} )} \left(\sum_{j = 0}^{m-1} \left[(\bzetat_i^{j+\frac{1}{2}})^2 + (\bzetat_i^{j+1})^2 - (\bzetat_i^{\prime j+\frac{1}{2}})^2 - (\bzetat_i^{\prime j+1})^2\right]\right.\right.\\
& \ \ \ \ \ \ \ \ \ \ \ \ \ \ \ \ \ \ \ \ \ \ \ \ \ \ \ \ \ \ \ \ \ \ \ \ \ \ \left.\left.+ \sum_{j=m}^{M-1} \left[(\bzeta_i^{j+\frac{1}{2}})^2 + (\bzeta_i^{j+1})^2 - (\bzeta_i^{\prime j+\frac{1}{2}})^2 - (\bzeta_i^{\prime j+1})^2\right]\right) \right],
\label{eq:genratio}
\end{align}
\end{widetext}
where \(m_i\) is the mass of particle \(i\).
Though this ratio appears to be cumbersome, it is straightforward to compute in terms of all of the noise variables \(\bxi\) and \(\bxit\), and it is then used in the acceptance probabilities of Eqs.~\eqref{eq:pacc} and~\eqref{eq:pacc2}.

Those acceptance probabilities can be further simplified by recognizing a cancellation in the product of Eq.~\eqref{eq:genratio} and the \(P(\xvec'|\blam',\mathbf{x}'(0)) / P(\xvec|\blam, \mathbf{x}(0))\) term of Eqs.~\eqref{eq:rhoratio} and~\eqref{eq:rhoratio2}.
For example, after some algebra, Eq.~\eqref{eq:pacc} becomes

\begin{widetext}
\begin{align}
\nonumber P_{\rm acc} & = \min \left[1, \ h_A(\xb'(0)) h_B(\xb'(\tobs)) \frac{Z_A(\blam')^{-1}} {Z_A(\blam)^{-1}} \frac{e^{-\beta H(\xb'(0);\blam')}}{e^{-\beta H(\xb(0);\blam)}}\right.
\\ & \left.\times \exp{ \left[ \sum_{i=1}^N \frac{\beta}{2 m_{i} \left(1 - e^{- \frac{\gamma \Delta t}{m_{i}}} \right)} \sum_{j=0}^{m-1} \left[ ( \bxi_{i}^{j+\frac{1}{2}} )^{2} + ( \bxi_{i}^{j+1} )^{2} - ( \bxi_{i}^{\prime j+\frac{1}{2}} )^{2} - ( \bxi_{i}^{\prime j+1} )^{2} + ( \bxit_{i}^{\prime j+\frac{1}{2}} )^{2} + ( \bxit_{i}^{\prime j+1} )^{2} - ( \bxit_{i}^{ j+\frac{1}{2}} )^{2} - ( \bxit_{i}^{ j+1} )^{2} \right] \right] } \right].
\label{eq:EqB4}
\end{align}
\end{widetext}

After the algebraic simplification, only contributions from the first \(m-1\) steps of the trajectory remain in Eq.~\eqref{eq:EqB4}.
Even this simplified expression looks daunting, but is easily computed by storing the random numbers of the current and trial trajectories.
The long sum of squares of \(\bxi\) terms has the physical interpretation of a heat flow between system and thermostat~\cite{athenes2008measurement}.
We want to weigh the trajectories based on their probability of occurring from forward-time integration, but we generated a portion of the trajectory (\(j = 0\) to \(j = m-1\)) from reversed-time integration.
To compute that trajectory's likelihood in the forward-time trajectory ensemble, we must reweight by an exponential of the heat.
We also note that because trial trajectories are always generated from previous successful trajectories, the terms \(h_{A}(\mathbf{x}(0))\) and \(h_{B}(\mathbf{x}(\tobs))\) are always 1.

For the simple Lennard-Jones binding system of Sections~\ref{sec:traj} and~\ref{sec:multi}, we used \(\alpha = 0.99\) and \(M=1000\) with a time step of \(\Delta t = 0.005\) for an observation time of \(\tobs = M \Delta t = 5\). 
For the particle escaping from a tetrahedral cage of Section~\ref{sec:complex}, we used \(\alpha = 0.9\) and \(M=1000\), again, with a time step of \(\Delta t = 0.005\) for an observation time of \(\tobs = 5\).
For both systems we built an initial trajectory by interpolating a starting state at the potential energy minima with Boltzmann momenta and an ending state just inside our respective definitions of the product state.
To build a starting trajectory one could run natural dynamics with an integrator until a suitable trajectory is isolated or do linear interpolation between a starting and ending point, as we have done.  
This linear interpolation leads to an unphysical starting trajectory, but sampling quickly moves away from it to more natural trajectories.

When sampling with multiple independent trajectories in order to bias more severely towards faster rates, our overall acceptance ratio is a product of individual ratios with the form of Eq.~\eqref{eq:EqB4}
For example, when the exact partition function is known, the acceptance probability using \(L\) trajectories \(\xvec_l\) with \(l = 1, 2, \ldots, L\) is:
\begin{widetext}
\begin{align}
\nonumber P_{\rm acc} & = \min \left[1, \  \frac{Z_A(\blam')^{-L}} {Z_A(\blam)^{-L}} \prod_{l=1}^{L} h_A(\xb'_{l}(0)) h_B(\xb'_{l}(\tobs)) \frac{e^{-\beta H(\xb'_{l}(0);\blam')}}{e^{-\beta H(\xb_{l}(0);\blam)}}\right.
\\ & \left.\times \exp{\left[ \sum_{i=1}^N \frac{\beta}{2 m_{i} \left(1 - e^{- \frac{\gamma \Delta t}{m_{i}}} \right) } \sum_{j=0}^{m-1} \left[ ( \bxi_{i,l}^{j+\frac{1}{2}} )^{2} + ( \bxi_{i,l}^{j+1} )^{2} - ( \bxi_{i,l}^{\prime j+\frac{1}{2}} )^{2} - ( \bxi_{i,l}^{\prime j+1} )^{2} + ( \bxit_{i,l}^{\prime j+\frac{1}{2}} )^{2} + ( \bxit_{i,l}^{\prime j+1} )^{2} - ( \bxit_{i,l}^{ j+\frac{1}{2}} )^{2} - ( \bxit_{i,l}^{ j+1} )^{2} \right] \right]} \right].
\label{eq:EqB5}
\end{align}
\end{widetext}
 When those partition functions are not known, they must be estimated as discussed in Section~\ref{sec:multitraj}.
In either case, the \(L\) trajectories share the same length and parameter set \(\blam\), but they vary in their independent starting configurations and in their noises \(\bxi_l\).

\section{Roulette procedure for producing unbiased series estimates} 
\label{sec:unbiased}
To obtain unbiased estimates of the partition function we must truncate the infinite series
\begin{equation}
S=1 + a^{(1)}(1 + a^{(2)}(1 + \ldots)) = 1 + \sum_{i=1}^{\infty} \prod_{j=1}^{i}a^{(j)}
 \label{eq:EqA1}
 \end{equation}
 using a roulette procedure, adapted from Ref.~\cite{booth2007unbiased}.
We introduce a roulette parameter \(R\) and calculate individual values of \(a^{(n)}\) sequentially as described in Section~\ref{sec:recipest}.
For the \(n^{\rm th}\) term in the expansion, if the running product \(\Pi^{(n)}\) of Eq.~\eqref{eq:runningproduct} is less than the parameter \(R\) then a roulette game is played with the series.  
 The series will continue with survival probability \(q^{(n)} = \Pi^{(n)}/R\) and truncate with probability \(1-q^{(n)}\). 
 If the series continues then we scale \(a^{(n)}\) by the survival probability \(q^{(n)}\) and continue the procedure for sample \(n+1\).
 If the series truncates then \(a^{(n)} \to 0\) and the estimate of the series is complete.  
Alternatively, when the running product is not less than \(R\), the series continues without scaling \(a^{(n)}\).
We compactly express the various cases by noting that sample \(n\) survives with probability \(q^{(n)} = \min[1, \Pi^{(n)}/R]\) in which case it contributes the scaled contribution \(a^{(n)} \max[1, R/\Pi^{(n)}]\).

As stated in the main text, the threshold for stochastic truncation after term \(n\), \(\Pi^{(n)}\), is computed recursively as the running product of the absolute value of these scaled contributions:
\begin{equation}
\Pi^{(n)} = \left| a^{(n)} \right| \prod_{i=1}^{n-1} \left| a^{(i)} \right|  \max\left[1, \frac{R}{\Pi^{(i)}}\right],
\label{eq:EqA3}
\end{equation}
where \(\Pi^{(1)} =  \left| a^{(1)} \right|\).
Since \(a^{(n)}\) values can be scaled throughout the procedure and because \(a^{(n)} = 0\) if the series is truncated at the \(n^{\rm th}\) term, the infinite series~\eqref{eq:EqA1} is replaced by
\begin{equation}
  S^{(n)} = 1 + \sum_{i=1}^{n-1} \prod_{j=1}^{i} a^{(j)} \max\left[1,\frac{R}{\Pi^{(j)}}\right].
  \label{eq:EqA5}
\end{equation}
 
We now show that the scaling of terms, which up to now was introduced in an \textit{ad hoc} manner, was constructed such that the expectation value of the truncated sum will exactly equal the expected value of the infinite sum.
Note that the expected value of the truncated series can be expressed as the sum over \(n\) of the probability of reaching the \(n^{\rm th}\) term times the value of the truncated sum \(S^{(n)}\).
The algebra of that expectation value simplifies due to telescoping sums to give
\begin{widetext}
\begin{align}
\nonumber 
\langle S \rangle &= \left(1 - q^{(1)}\right) S^{(1)} + \left(1 - q^{(2)}\right) q^{(1)} S^{(2)} + \left(1 - q^{(3)}\right) q^{(2)} q^{(1)} S^{(3)} + \hdots\\
\nonumber &= \left(1 - q^{(1)}\right) + \left(1 - q^{(2)}\right)q^{(1)}\left(1 + a^{(1)} \max\left[1, \frac{R}{\Pi^{(1)}}\right]\right)\\
\nonumber & \ \ \ \  + \left(1 - q^{(3)}\right)q^{(1)}q^{(2)}\left(1 + a^{(1)} \max\left[1, \frac{R}{\Pi^{(1)}}\right] + a^{(1)}a^{(2)} \max \left[1,\frac{R}{\Pi^{(1)}}\right]\max \left[1,\frac{R}{\Pi^{(2)}}\right]\right) + \hdots\\
\nonumber &= 1 + q^{(1)}a^{(1)} \max\left[1,\frac{R}{\Pi^{(1)}}\right] + q^{(1)} q^{(2)} a^{(1)} a^{(2)} \max\left[1, \frac{R}{\Pi^{(1)}}\right] \max\left[1, \frac{R}{\Pi^{(2)}}\right] + \hdots\\
\nonumber &= 1 + \sum_{i=1}^\infty \prod_{j=1}^i a^{(j)} \min\left[1,\frac{\Pi^{(j)}}{R}\right] \max\left[1, \frac{R}{\Pi^{(j)}}\right]\\
&=  1 + \sum_{i=1}^\infty \prod_{j=1}^i a^{(j)}
 \label{eq:EqA6}
\end{align}
\end{widetext}
The final equality follows because \(R > 0\), \(\Pi^{(i)} > 0\), and for positive \(x\), \(\min[1,x]\max[1,x^{-1}] = 1\).
Consequently, we see that the expectation value of the truncated sums, Eq.~\eqref{eq:EqA6}, equals that of the infinite sum, Eq.~\eqref{eq:EqA1}.
In other words, the stochastic truncation scheme is unbiased.
We note that the roulette procedure we have described is not a unique way to generate an unbiased stochastic truncation.
It may be possible to design alternative roulette games which give a better trade-off between truncation speed and estimate noise.

\bibliography{biblio.bib}

\begin{thebibliography}{47}%
\makeatletter
\providecommand \@ifxundefined [1]{%
 \@ifx{#1\undefined}
}%
\providecommand \@ifnum [1]{%
 \ifnum #1\expandafter \@firstoftwo
 \else \expandafter \@secondoftwo
 \fi
}%
\providecommand \@ifx [1]{%
 \ifx #1\expandafter \@firstoftwo
 \else \expandafter \@secondoftwo
 \fi
}%
\providecommand \natexlab [1]{#1}%
\providecommand \enquote  [1]{``#1''}%
\providecommand \bibnamefont  [1]{#1}%
\providecommand \bibfnamefont [1]{#1}%
\providecommand \citenamefont [1]{#1}%
\providecommand \href@noop [0]{\@secondoftwo}%
\providecommand \href [0]{\begingroup \@sanitize@url \@href}%
\providecommand \@href[1]{\@@startlink{#1}\@@href}%
\providecommand \@@href[1]{\endgroup#1\@@endlink}%
\providecommand \@sanitize@url [0]{\catcode `\\12\catcode `\$12\catcode
  `\&12\catcode `\#12\catcode `\^12\catcode `\_12\catcode `\%12\relax}%
\providecommand \@@startlink[1]{}%
\providecommand \@@endlink[0]{}%
\providecommand \url  [0]{\begingroup\@sanitize@url \@url }%
\providecommand \@url [1]{\endgroup\@href {#1}{\urlprefix }}%
\providecommand \urlprefix  [0]{URL }%
\providecommand \Eprint [0]{\href }%
\providecommand \doibase [0]{http://dx.doi.org/}%
\providecommand \selectlanguage [0]{\@gobble}%
\providecommand \bibinfo  [0]{\@secondoftwo}%
\providecommand \bibfield  [0]{\@secondoftwo}%
\providecommand \translation [1]{[#1]}%
\providecommand \BibitemOpen [0]{}%
\providecommand \bibitemStop [0]{}%
\providecommand \bibitemNoStop [0]{.\EOS\space}%
\providecommand \EOS [0]{\spacefactor3000\relax}%
\providecommand \BibitemShut  [1]{\csname bibitem#1\endcsname}%
\let\auto@bib@innerbib\@empty
\bibitem [{\citenamefont {Kramers}(1940)}]{kramers1940brownian}%
  \BibitemOpen
  \bibfield  {author} {\bibinfo {author} {\bibfnamefont {H.~A.}\ \bibnamefont
  {Kramers}},\ }\href {\doibase 10.1016/S0031-8914(40)90098-2} {\bibfield
  {journal} {\bibinfo  {journal} {Physica}\ }\textbf {\bibinfo {volume} {7}},\
  \bibinfo {pages} {284 } (\bibinfo {year} {1940})}\BibitemShut {NoStop}%
\bibitem [{\citenamefont {H{\"a}nggi}\ \emph {et~al.}(1990)\citenamefont
  {H{\"a}nggi}, \citenamefont {Talkner},\ and\ \citenamefont
  {Borkovec}}]{hanggi1990reaction}%
  \BibitemOpen
  \bibfield  {author} {\bibinfo {author} {\bibfnamefont {P.}~\bibnamefont
  {H{\"a}nggi}}, \bibinfo {author} {\bibfnamefont {P.}~\bibnamefont {Talkner}},
  \ and\ \bibinfo {author} {\bibfnamefont {M.}~\bibnamefont {Borkovec}},\
  }\href {\doibase 10.1103/RevModPhys.62.251} {\bibfield  {journal} {\bibinfo
  {journal} {Reviews of Modern Physics}\ }\textbf {\bibinfo {volume} {62}},\
  \bibinfo {pages} {251} (\bibinfo {year} {1990})}\BibitemShut {NoStop}%
\bibitem [{\citenamefont {Bolhuis}\ \emph {et~al.}(2002)\citenamefont
  {Bolhuis}, \citenamefont {Chandler}, \citenamefont {Dellago},\ and\
  \citenamefont {Geissler}}]{bolhuis2002transition}%
  \BibitemOpen
  \bibfield  {author} {\bibinfo {author} {\bibfnamefont {P.~G.}\ \bibnamefont
  {Bolhuis}}, \bibinfo {author} {\bibfnamefont {D.}~\bibnamefont {Chandler}},
  \bibinfo {author} {\bibfnamefont {C.}~\bibnamefont {Dellago}}, \ and\
  \bibinfo {author} {\bibfnamefont {P.~L.}\ \bibnamefont {Geissler}},\ }\href
  {\doibase 10.1146/annurev.physchem.53.082301.113146} {\bibfield  {journal}
  {\bibinfo  {journal} {Annual Review of Physical Chemistry}\ }\textbf
  {\bibinfo {volume} {53}},\ \bibinfo {pages} {291} (\bibinfo {year}
  {2002})}\BibitemShut {NoStop}%
\bibitem [{\citenamefont {Peters}(2010)}]{peters2010recent}%
  \BibitemOpen
  \bibfield  {author} {\bibinfo {author} {\bibfnamefont {B.}~\bibnamefont
  {Peters}},\ }\href {\doibase 10.1080/08927020903536382} {\bibfield  {journal}
  {\bibinfo  {journal} {Molecular Simulation}\ }\textbf {\bibinfo {volume}
  {36}},\ \bibinfo {pages} {1265} (\bibinfo {year} {2010})}\BibitemShut
  {NoStop}%
\bibitem [{\citenamefont {Yamamoto}(1960)}]{yamamoto1960quantum}%
  \BibitemOpen
  \bibfield  {author} {\bibinfo {author} {\bibfnamefont {T.}~\bibnamefont
  {Yamamoto}},\ }\href {\doibase 10.1063/1.1731099} {\bibfield  {journal}
  {\bibinfo  {journal} {The Journal of Chemical Physics}\ }\textbf {\bibinfo
  {volume} {33}},\ \bibinfo {pages} {281} (\bibinfo {year} {1960})}\BibitemShut
  {NoStop}%
\bibitem [{\citenamefont {Miller}\ \emph {et~al.}(1983)\citenamefont {Miller},
  \citenamefont {Schwartz},\ and\ \citenamefont {Tromp}}]{miller1983quantum}%
  \BibitemOpen
  \bibfield  {author} {\bibinfo {author} {\bibfnamefont {W.~H.}\ \bibnamefont
  {Miller}}, \bibinfo {author} {\bibfnamefont {S.~D.}\ \bibnamefont
  {Schwartz}}, \ and\ \bibinfo {author} {\bibfnamefont {J.~W.}\ \bibnamefont
  {Tromp}},\ }\href {\doibase https://doi.org/10.1063/1.445581} {\bibfield
  {journal} {\bibinfo  {journal} {The Journal of Chemical Physics}\ }\textbf
  {\bibinfo {volume} {79}},\ \bibinfo {pages} {4889} (\bibinfo {year}
  {1983})}\BibitemShut {NoStop}%
\bibitem [{\citenamefont {Chandler}(1978)}]{chandler1978statistical}%
  \BibitemOpen
  \bibfield  {author} {\bibinfo {author} {\bibfnamefont {D.}~\bibnamefont
  {Chandler}},\ }\href {\doibase 10.1063/1.436049} {\bibfield  {journal}
  {\bibinfo  {journal} {The Journal of Chemical Physics}\ }\textbf {\bibinfo
  {volume} {68}},\ \bibinfo {pages} {2959} (\bibinfo {year}
  {1978})}\BibitemShut {NoStop}%
\bibitem [{\citenamefont {Montgomery~Jr.}\ \emph {et~al.}(1979)\citenamefont
  {Montgomery~Jr.}, \citenamefont {Chandler},\ and\ \citenamefont
  {Berne}}]{montgomery1979trajectory}%
  \BibitemOpen
  \bibfield  {author} {\bibinfo {author} {\bibfnamefont {J.~A.}\ \bibnamefont
  {Montgomery~Jr.}}, \bibinfo {author} {\bibfnamefont {D.}~\bibnamefont
  {Chandler}}, \ and\ \bibinfo {author} {\bibfnamefont {B.~J.}\ \bibnamefont
  {Berne}},\ }\href {\doibase 10.1063/1.438028} {\bibfield  {journal} {\bibinfo
   {journal} {The Journal of Chemical Physics}\ }\textbf {\bibinfo {volume}
  {70}},\ \bibinfo {pages} {4056} (\bibinfo {year} {1979})}\BibitemShut
  {NoStop}%
\bibitem [{\citenamefont {Chandler}(1986)}]{chandler1986roles}%
  \BibitemOpen
  \bibfield  {author} {\bibinfo {author} {\bibfnamefont {D.}~\bibnamefont
  {Chandler}},\ }\href {\doibase 10.1007/BF01010840} {\bibfield  {journal}
  {\bibinfo  {journal} {Journal of Statistical Physics}\ }\textbf {\bibinfo
  {volume} {42}},\ \bibinfo {pages} {49} (\bibinfo {year} {1986})}\BibitemShut
  {NoStop}%
\bibitem [{\citenamefont {Chandler}\ and\ \citenamefont
  {Kuharski}(1988)}]{chandler1988two}%
  \BibitemOpen
  \bibfield  {author} {\bibinfo {author} {\bibfnamefont {D.}~\bibnamefont
  {Chandler}}\ and\ \bibinfo {author} {\bibfnamefont {R.~A.}\ \bibnamefont
  {Kuharski}},\ }\href {\doibase 10.1039/DC9888500329} {\bibfield  {journal}
  {\bibinfo  {journal} {Faraday Discussions of the Chemical Society}\ }\textbf
  {\bibinfo {volume} {85}},\ \bibinfo {pages} {329} (\bibinfo {year}
  {1988})}\BibitemShut {NoStop}%
\bibitem [{\citenamefont {Borkovec}\ and\ \citenamefont
  {Talkner}(1990)}]{borkovec1990generalized}%
  \BibitemOpen
  \bibfield  {author} {\bibinfo {author} {\bibfnamefont {M.}~\bibnamefont
  {Borkovec}}\ and\ \bibinfo {author} {\bibfnamefont {P.}~\bibnamefont
  {Talkner}},\ }\href {\doibase 10.1063/1.458535} {\bibfield  {journal}
  {\bibinfo  {journal} {The Journal of Chemical Physics}\ }\textbf {\bibinfo
  {volume} {92}},\ \bibinfo {pages} {5307} (\bibinfo {year}
  {1990})}\BibitemShut {NoStop}%
\bibitem [{\citenamefont {Hummer}(2004)}]{hummer2004transition}%
  \BibitemOpen
  \bibfield  {author} {\bibinfo {author} {\bibfnamefont {G.}~\bibnamefont
  {Hummer}},\ }\href {\doibase 10.1063/1.1630572} {\bibfield  {journal}
  {\bibinfo  {journal} {The Journal of Chemical Physics}\ }\textbf {\bibinfo
  {volume} {120}},\ \bibinfo {pages} {516} (\bibinfo {year}
  {2004})}\BibitemShut {NoStop}%
\bibitem [{\citenamefont {Pratt}(1986)}]{pratt1986statistical}%
  \BibitemOpen
  \bibfield  {author} {\bibinfo {author} {\bibfnamefont {L.~R.}\ \bibnamefont
  {Pratt}},\ }\href {\doibase 10.1063/1.451695} {\bibfield  {journal} {\bibinfo
   {journal} {The Journal of Chemical Physics}\ }\textbf {\bibinfo {volume}
  {85}},\ \bibinfo {pages} {5045} (\bibinfo {year} {1986})}\BibitemShut
  {NoStop}%
\bibitem [{\citenamefont {Dellago}\ \emph {et~al.}(2002)\citenamefont
  {Dellago}, \citenamefont {Bolhuis},\ and\ \citenamefont
  {Geissler}}]{dellago2002transition}%
  \BibitemOpen
  \bibfield  {author} {\bibinfo {author} {\bibfnamefont {C.}~\bibnamefont
  {Dellago}}, \bibinfo {author} {\bibfnamefont {P.}~\bibnamefont {Bolhuis}}, \
  and\ \bibinfo {author} {\bibfnamefont {P.~L.}\ \bibnamefont {Geissler}},\
  }\href {\doibase 10.1002/0471231509.ch1} {\bibfield  {journal} {\bibinfo
  {journal} {Advances in Chemical Physics}\ }\textbf {\bibinfo {volume}
  {123}},\ \bibinfo {pages} {1} (\bibinfo {year} {2002})}\BibitemShut {NoStop}%
\bibitem [{\citenamefont {Bolhuis}(2002)}]{bolhuis2002transition2}%
  \BibitemOpen
  \bibfield  {author} {\bibinfo {author} {\bibfnamefont {P.~G.}\ \bibnamefont
  {Bolhuis}},\ }\href {\doibase 10.1088/0953-8984/15/1/314} {\bibfield
  {journal} {\bibinfo  {journal} {Journal of Physics: Condensed Matter}\
  }\textbf {\bibinfo {volume} {15}},\ \bibinfo {pages} {S113} (\bibinfo {year}
  {2002})}\BibitemShut {NoStop}%
\bibitem [{\citenamefont {Dellago}\ \emph {et~al.}(1998)\citenamefont
  {Dellago}, \citenamefont {Bolhuis}, \citenamefont {Csajka},\ and\
  \citenamefont {Chandler}}]{dellago1998transition}%
  \BibitemOpen
  \bibfield  {author} {\bibinfo {author} {\bibfnamefont {C.}~\bibnamefont
  {Dellago}}, \bibinfo {author} {\bibfnamefont {P.~G.}\ \bibnamefont
  {Bolhuis}}, \bibinfo {author} {\bibfnamefont {F.~S.}\ \bibnamefont {Csajka}},
  \ and\ \bibinfo {author} {\bibfnamefont {D.}~\bibnamefont {Chandler}},\
  }\href {\doibase 10.1063/1.475562} {\bibfield  {journal} {\bibinfo  {journal}
  {The Journal of Chemical Physics}\ }\textbf {\bibinfo {volume} {108}},\
  \bibinfo {pages} {1964} (\bibinfo {year} {1998})}\BibitemShut {NoStop}%
\bibitem [{\citenamefont {G.~Bolhuis}\ \emph {et~al.}(1998)\citenamefont
  {G.~Bolhuis}, \citenamefont {Dellago},\ and\ \citenamefont
  {Chandler}}]{bolhuis1998sampling}%
  \BibitemOpen
  \bibfield  {author} {\bibinfo {author} {\bibfnamefont {P.}~\bibnamefont
  {G.~Bolhuis}}, \bibinfo {author} {\bibfnamefont {C.}~\bibnamefont {Dellago}},
  \ and\ \bibinfo {author} {\bibfnamefont {D.}~\bibnamefont {Chandler}},\
  }\href {\doibase 10.1039/A801266K} {\bibfield  {journal} {\bibinfo  {journal}
  {Faraday Discuss.}\ }\textbf {\bibinfo {volume} {110}},\ \bibinfo {pages}
  {421} (\bibinfo {year} {1998})}\BibitemShut {NoStop}%
\bibitem [{\citenamefont {Dellago}\ \emph {et~al.}(1999)\citenamefont
  {Dellago}, \citenamefont {Bolhuis},\ and\ \citenamefont
  {Chandler}}]{dellago1999calculation}%
  \BibitemOpen
  \bibfield  {author} {\bibinfo {author} {\bibfnamefont {C.}~\bibnamefont
  {Dellago}}, \bibinfo {author} {\bibfnamefont {P.~G.}\ \bibnamefont
  {Bolhuis}}, \ and\ \bibinfo {author} {\bibfnamefont {D.}~\bibnamefont
  {Chandler}},\ }\href {\doibase 10.1063/1.478569} {\bibfield  {journal}
  {\bibinfo  {journal} {The Journal of Chemical Physics}\ }\textbf {\bibinfo
  {volume} {110}},\ \bibinfo {pages} {6617} (\bibinfo {year}
  {1999})}\BibitemShut {NoStop}%
\bibitem [{\citenamefont {Peters}\ and\ \citenamefont
  {Trout}(2006)}]{peters2006obtaining}%
  \BibitemOpen
  \bibfield  {author} {\bibinfo {author} {\bibfnamefont {B.}~\bibnamefont
  {Peters}}\ and\ \bibinfo {author} {\bibfnamefont {B.~L.}\ \bibnamefont
  {Trout}},\ }\href {\doibase 10.1063/1.2234477} {\bibfield  {journal}
  {\bibinfo  {journal} {The Journal of Chemical Physics}\ }\textbf {\bibinfo
  {volume} {125}},\ \bibinfo {pages} {054108} (\bibinfo {year}
  {2006})}\BibitemShut {NoStop}%
\bibitem [{\citenamefont {Peters}\ \emph {et~al.}(2007)\citenamefont {Peters},
  \citenamefont {Beckham},\ and\ \citenamefont {Trout}}]{peters2007extensions}%
  \BibitemOpen
  \bibfield  {author} {\bibinfo {author} {\bibfnamefont {B.}~\bibnamefont
  {Peters}}, \bibinfo {author} {\bibfnamefont {G.~T.}\ \bibnamefont {Beckham}},
  \ and\ \bibinfo {author} {\bibfnamefont {B.~L.}\ \bibnamefont {Trout}},\
  }\href {\doibase 10.1063/1.2748396} {\bibfield  {journal} {\bibinfo
  {journal} {The Journal of Chemical Physics}\ }\textbf {\bibinfo {volume}
  {127}},\ \bibinfo {pages} {034109} (\bibinfo {year} {2007})}\BibitemShut
  {NoStop}%
\bibitem [{\citenamefont {Miller~III}\ and\ \citenamefont
  {Predescu}(2007)}]{miller2007sampling}%
  \BibitemOpen
  \bibfield  {author} {\bibinfo {author} {\bibfnamefont {T.~F.}\ \bibnamefont
  {Miller~III}}\ and\ \bibinfo {author} {\bibfnamefont {C.}~\bibnamefont
  {Predescu}},\ }\href {\doibase 10.1063/1.2712444} {\bibfield  {journal}
  {\bibinfo  {journal} {The Journal of Chemical Physics}\ }\textbf {\bibinfo
  {volume} {126}},\ \bibinfo {pages} {144102} (\bibinfo {year}
  {2007})}\BibitemShut {NoStop}%
\bibitem [{\citenamefont {Gr{\"u}nwald}\ \emph {et~al.}(2008)\citenamefont
  {Gr{\"u}nwald}, \citenamefont {Dellago},\ and\ \citenamefont
  {Geissler}}]{grunwald2008precision}%
  \BibitemOpen
  \bibfield  {author} {\bibinfo {author} {\bibfnamefont {M.}~\bibnamefont
  {Gr{\"u}nwald}}, \bibinfo {author} {\bibfnamefont {C.}~\bibnamefont
  {Dellago}}, \ and\ \bibinfo {author} {\bibfnamefont {P.~L.}\ \bibnamefont
  {Geissler}},\ }\href {\doibase 10.1063/1.2978000} {\bibfield  {journal}
  {\bibinfo  {journal} {The Journal of Chemical Physics}\ }\textbf {\bibinfo
  {volume} {129}},\ \bibinfo {pages} {194101} (\bibinfo {year}
  {2008})}\BibitemShut {NoStop}%
\bibitem [{\citenamefont {van Erp}\ \emph {et~al.}(2003)\citenamefont {van
  Erp}, \citenamefont {Moroni},\ and\ \citenamefont {Bolhuis}}]{van2003novel}%
  \BibitemOpen
  \bibfield  {author} {\bibinfo {author} {\bibfnamefont {T.~S.}\ \bibnamefont
  {van Erp}}, \bibinfo {author} {\bibfnamefont {D.}~\bibnamefont {Moroni}}, \
  and\ \bibinfo {author} {\bibfnamefont {P.~G.}\ \bibnamefont {Bolhuis}},\
  }\href {\doibase 10.1063/1.1562614} {\bibfield  {journal} {\bibinfo
  {journal} {The Journal of Chemical Physics}\ }\textbf {\bibinfo {volume}
  {118}},\ \bibinfo {pages} {7762} (\bibinfo {year} {2003})}\BibitemShut
  {NoStop}%
\bibitem [{\citenamefont {Moroni}\ \emph {et~al.}(2004)\citenamefont {Moroni},
  \citenamefont {Bolhuis},\ and\ \citenamefont {van Erp}}]{moroni2004rate}%
  \BibitemOpen
  \bibfield  {author} {\bibinfo {author} {\bibfnamefont {D.}~\bibnamefont
  {Moroni}}, \bibinfo {author} {\bibfnamefont {P.~G.}\ \bibnamefont {Bolhuis}},
  \ and\ \bibinfo {author} {\bibfnamefont {T.~S.}\ \bibnamefont {van Erp}},\
  }\href {\doibase 10.1063/1.1644537} {\bibfield  {journal} {\bibinfo
  {journal} {The Journal of Chemical Physics}\ }\textbf {\bibinfo {volume}
  {120}},\ \bibinfo {pages} {4055} (\bibinfo {year} {2004})}\BibitemShut
  {NoStop}%
\bibitem [{\citenamefont {van Erp}\ and\ \citenamefont
  {Bolhuis}(2005)}]{van2005elaborating}%
  \BibitemOpen
  \bibfield  {author} {\bibinfo {author} {\bibfnamefont {T.~S.}\ \bibnamefont
  {van Erp}}\ and\ \bibinfo {author} {\bibfnamefont {P.~G.}\ \bibnamefont
  {Bolhuis}},\ }\href {\doibase 10.1016/j.jcp.2004.11.003} {\bibfield
  {journal} {\bibinfo  {journal} {Journal of Computational Physics}\ }\textbf
  {\bibinfo {volume} {205}},\ \bibinfo {pages} {157} (\bibinfo {year}
  {2005})}\BibitemShut {NoStop}%
\bibitem [{\citenamefont {Allen}\ \emph {et~al.}(2005)\citenamefont {Allen},
  \citenamefont {Warren},\ and\ \citenamefont {ten Wolde}}]{allen2005sampling}%
  \BibitemOpen
  \bibfield  {author} {\bibinfo {author} {\bibfnamefont {R.~J.}\ \bibnamefont
  {Allen}}, \bibinfo {author} {\bibfnamefont {P.~B.}\ \bibnamefont {Warren}}, \
  and\ \bibinfo {author} {\bibfnamefont {P.~R.}\ \bibnamefont {ten Wolde}},\
  }\href {\doibase 10.1103/PhysRevLett.94.018104} {\bibfield  {journal}
  {\bibinfo  {journal} {Physical review letters}\ }\textbf {\bibinfo {volume}
  {94}},\ \bibinfo {pages} {018104} (\bibinfo {year} {2005})}\BibitemShut
  {NoStop}%
\bibitem [{\citenamefont {Allen}\ \emph
  {et~al.}(2006{\natexlab{a}})\citenamefont {Allen}, \citenamefont {Frenkel},\
  and\ \citenamefont {ten Wolde}}]{allen2006simulating}%
  \BibitemOpen
  \bibfield  {author} {\bibinfo {author} {\bibfnamefont {R.~J.}\ \bibnamefont
  {Allen}}, \bibinfo {author} {\bibfnamefont {D.}~\bibnamefont {Frenkel}}, \
  and\ \bibinfo {author} {\bibfnamefont {P.~R.}\ \bibnamefont {ten Wolde}},\
  }\href {\doibase 10.1063/1.2140273} {\bibfield  {journal} {\bibinfo
  {journal} {The Journal of Chemical Physics}\ }\textbf {\bibinfo {volume}
  {124}},\ \bibinfo {pages} {024102} (\bibinfo {year}
  {2006}{\natexlab{a}})}\BibitemShut {NoStop}%
\bibitem [{\citenamefont {Allen}\ \emph
  {et~al.}(2006{\natexlab{b}})\citenamefont {Allen}, \citenamefont {Frenkel},\
  and\ \citenamefont {ten Wolde}}]{allen2006forward}%
  \BibitemOpen
  \bibfield  {author} {\bibinfo {author} {\bibfnamefont {R.~J.}\ \bibnamefont
  {Allen}}, \bibinfo {author} {\bibfnamefont {D.}~\bibnamefont {Frenkel}}, \
  and\ \bibinfo {author} {\bibfnamefont {P.~R.}\ \bibnamefont {ten Wolde}},\
  }\href {\doibase 10.1063/1.2198827} {\bibfield  {journal} {\bibinfo
  {journal} {The Journal of Chemical Physics}\ }\textbf {\bibinfo {volume}
  {124}},\ \bibinfo {pages} {194111} (\bibinfo {year}
  {2006}{\natexlab{b}})}\BibitemShut {NoStop}%
\bibitem [{\citenamefont {Allen}\ \emph {et~al.}(2009)\citenamefont {Allen},
  \citenamefont {Valeriani},\ and\ \citenamefont {ten
  Wolde}}]{allen2009forward}%
  \BibitemOpen
  \bibfield  {author} {\bibinfo {author} {\bibfnamefont {R.~J.}\ \bibnamefont
  {Allen}}, \bibinfo {author} {\bibfnamefont {C.}~\bibnamefont {Valeriani}}, \
  and\ \bibinfo {author} {\bibfnamefont {P.~R.}\ \bibnamefont {ten Wolde}},\
  }\href {\doibase 10.1088/0953-8984/21/46/463102} {\bibfield  {journal}
  {\bibinfo  {journal} {Journal of Physics: Condensed Matter}\ }\textbf
  {\bibinfo {volume} {21}},\ \bibinfo {pages} {463102} (\bibinfo {year}
  {2009})}\BibitemShut {NoStop}%
\bibitem [{\citenamefont {Borrero}\ and\ \citenamefont
  {Escobedo}(2008)}]{borrero2008optimizing}%
  \BibitemOpen
  \bibfield  {author} {\bibinfo {author} {\bibfnamefont {E.~E.}\ \bibnamefont
  {Borrero}}\ and\ \bibinfo {author} {\bibfnamefont {F.~A.}\ \bibnamefont
  {Escobedo}},\ }\href {\doibase 10.1063/1.2953325} {\bibfield  {journal}
  {\bibinfo  {journal} {The Journal of Chemical Physics}\ }\textbf {\bibinfo
  {volume} {129}},\ \bibinfo {pages} {024115} (\bibinfo {year}
  {2008})}\BibitemShut {NoStop}%
\bibitem [{\citenamefont {Borrero}\ and\ \citenamefont
  {Escobedo}(2009)}]{borrero2009simulating}%
  \BibitemOpen
  \bibfield  {author} {\bibinfo {author} {\bibfnamefont {E.~E.}\ \bibnamefont
  {Borrero}}\ and\ \bibinfo {author} {\bibfnamefont {F.~A.}\ \bibnamefont
  {Escobedo}},\ }\href {\doibase https://doi.org/10.1021/jp809103k} {\bibfield
  {journal} {\bibinfo  {journal} {The Journal of Physical Chemistry B}\
  }\textbf {\bibinfo {volume} {113}},\ \bibinfo {pages} {6434} (\bibinfo {year}
  {2009})}\BibitemShut {NoStop}%
\bibitem [{\citenamefont {Ceperley}\ and\ \citenamefont
  {Dewing}(1999)}]{ceperley1999penalty}%
  \BibitemOpen
  \bibfield  {author} {\bibinfo {author} {\bibfnamefont {D.~M.}\ \bibnamefont
  {Ceperley}}\ and\ \bibinfo {author} {\bibfnamefont {M.}~\bibnamefont
  {Dewing}},\ }\href {\doibase 10.1063/1.478034} {\bibfield  {journal}
  {\bibinfo  {journal} {The Journal of Chemical Physics}\ }\textbf {\bibinfo
  {volume} {110}},\ \bibinfo {pages} {9812} (\bibinfo {year}
  {1999})}\BibitemShut {NoStop}%
\bibitem [{\citenamefont {Booth}(2007)}]{booth2007unbiased}%
  \BibitemOpen
  \bibfield  {author} {\bibinfo {author} {\bibfnamefont {T.~E.}\ \bibnamefont
  {Booth}},\ }\href {\doibase 10.13182/NSE07-A2707} {\bibfield  {journal}
  {\bibinfo  {journal} {Nuclear Science and Engineering}\ }\textbf {\bibinfo
  {volume} {156}},\ \bibinfo {pages} {403} (\bibinfo {year}
  {2007})}\BibitemShut {NoStop}%
\bibitem [{\citenamefont {Gingrich}\ and\ \citenamefont
  {Geissler}(2015)}]{gingrich2015preserving}%
  \BibitemOpen
  \bibfield  {author} {\bibinfo {author} {\bibfnamefont {T.~R.}\ \bibnamefont
  {Gingrich}}\ and\ \bibinfo {author} {\bibfnamefont {P.~L.}\ \bibnamefont
  {Geissler}},\ }\href {\doibase 10.1063/1.4922343} {\bibfield  {journal}
  {\bibinfo  {journal} {The Journal of Chemical Physics}\ }\textbf {\bibinfo
  {volume} {142}},\ \bibinfo {pages} {234104} (\bibinfo {year}
  {2015})}\BibitemShut {NoStop}%
\bibitem [{\citenamefont {Diaconis}\ \emph {et~al.}(2000)\citenamefont
  {Diaconis}, \citenamefont {Holmes},\ and\ \citenamefont
  {Neal}}]{diaconis2000analysis}%
  \BibitemOpen
  \bibfield  {author} {\bibinfo {author} {\bibfnamefont {P.}~\bibnamefont
  {Diaconis}}, \bibinfo {author} {\bibfnamefont {S.}~\bibnamefont {Holmes}}, \
  and\ \bibinfo {author} {\bibfnamefont {R.~M.}\ \bibnamefont {Neal}},\ }\href
  {\doibase doi:10.1214/aoap/1019487508} {\bibfield  {journal} {\bibinfo
  {journal} {Annals of Applied Probability}\ }\textbf {\bibinfo {volume}
  {10}},\ \bibinfo {pages} {726} (\bibinfo {year} {2000})}\BibitemShut
  {NoStop}%
\bibitem [{\citenamefont {Lin}\ \emph {et~al.}(2000)\citenamefont {Lin},
  \citenamefont {Liu},\ and\ \citenamefont {Sloan}}]{lin2000noisy}%
  \BibitemOpen
  \bibfield  {author} {\bibinfo {author} {\bibfnamefont {L.}~\bibnamefont
  {Lin}}, \bibinfo {author} {\bibfnamefont {K.}~\bibnamefont {Liu}}, \ and\
  \bibinfo {author} {\bibfnamefont {J.}~\bibnamefont {Sloan}},\ }\href
  {\doibase 10.1103/PhysRevD.61.074505} {\bibfield  {journal} {\bibinfo
  {journal} {Physical Review D}\ }\textbf {\bibinfo {volume} {61}},\ \bibinfo
  {pages} {074505} (\bibinfo {year} {2000})}\BibitemShut {NoStop}%
\bibitem [{\citenamefont {Andrieu}\ and\ \citenamefont
  {Roberts}(2009)}]{andrieu2009pseudo}%
  \BibitemOpen
  \bibfield  {author} {\bibinfo {author} {\bibfnamefont {C.}~\bibnamefont
  {Andrieu}}\ and\ \bibinfo {author} {\bibfnamefont {G.~O.}\ \bibnamefont
  {Roberts}},\ }\href {\doibase 10.1214/07-AOS574} {\bibfield  {journal}
  {\bibinfo  {journal} {The Annals of Statistics}\ }\textbf {\bibinfo {volume}
  {37}},\ \bibinfo {pages} {697} (\bibinfo {year} {2009})}\BibitemShut
  {NoStop}%
\bibitem [{\citenamefont {Bernard}\ \emph {et~al.}(2009)\citenamefont
  {Bernard}, \citenamefont {Krauth},\ and\ \citenamefont
  {Wilson}}]{bernard2009event}%
  \BibitemOpen
  \bibfield  {author} {\bibinfo {author} {\bibfnamefont {E.~P.}\ \bibnamefont
  {Bernard}}, \bibinfo {author} {\bibfnamefont {W.}~\bibnamefont {Krauth}}, \
  and\ \bibinfo {author} {\bibfnamefont {D.~B.}\ \bibnamefont {Wilson}},\
  }\href {\doibase 10.1103/PhysRevE.80.056704} {\bibfield  {journal} {\bibinfo
  {journal} {Physical Review E}\ }\textbf {\bibinfo {volume} {80}},\ \bibinfo
  {pages} {056704} (\bibinfo {year} {2009})}\BibitemShut {NoStop}%
\bibitem [{\citenamefont {Vucelja}(2016)}]{vucelja2016lifting}%
  \BibitemOpen
  \bibfield  {author} {\bibinfo {author} {\bibfnamefont {M.}~\bibnamefont
  {Vucelja}},\ }\href {\doibase 10.1119/1.4961596} {\bibfield  {journal}
  {\bibinfo  {journal} {American Journal of Physics}\ }\textbf {\bibinfo
  {volume} {84}},\ \bibinfo {pages} {958} (\bibinfo {year} {2016})}\BibitemShut
  {NoStop}%
\bibitem [{\citenamefont {Lyne}\ \emph {et~al.}(2015)\citenamefont {Lyne},
  \citenamefont {Girolami}, \citenamefont {Atchad{\'e}}, \citenamefont
  {Strathmann},\ and\ \citenamefont {Simpson}}]{lyne2015russian}%
  \BibitemOpen
  \bibfield  {author} {\bibinfo {author} {\bibfnamefont {A.-M.}\ \bibnamefont
  {Lyne}}, \bibinfo {author} {\bibfnamefont {M.}~\bibnamefont {Girolami}},
  \bibinfo {author} {\bibfnamefont {Y.}~\bibnamefont {Atchad{\'e}}}, \bibinfo
  {author} {\bibfnamefont {H.}~\bibnamefont {Strathmann}}, \ and\ \bibinfo
  {author} {\bibfnamefont {D.}~\bibnamefont {Simpson}},\ }\href {\doibase
  10.1214/15-STS523} {\bibfield  {journal} {\bibinfo  {journal} {Statistical
  Science}\ }\textbf {\bibinfo {volume} {30}},\ \bibinfo {pages} {443}
  (\bibinfo {year} {2015})}\BibitemShut {NoStop}%
\bibitem [{\citenamefont {Bhanot}\ and\ \citenamefont
  {Kennedy}(1985)}]{bhanot1985bosonic}%
  \BibitemOpen
  \bibfield  {author} {\bibinfo {author} {\bibfnamefont {G.}~\bibnamefont
  {Bhanot}}\ and\ \bibinfo {author} {\bibfnamefont {A.~D.}\ \bibnamefont
  {Kennedy}},\ }\href {\doibase 10.1016/0370-2693(85)91214-6} {\bibfield
  {journal} {\bibinfo  {journal} {Physics Letters B}\ }\textbf {\bibinfo
  {volume} {157}},\ \bibinfo {pages} {70} (\bibinfo {year} {1985})}\BibitemShut
  {NoStop}%
\bibitem [{\citenamefont {Gingrich}\ \emph {et~al.}(2016)\citenamefont
  {Gingrich}, \citenamefont {Rotskoff}, \citenamefont {Crooks},\ and\
  \citenamefont {Geissler}}]{gingrich2016near}%
  \BibitemOpen
  \bibfield  {author} {\bibinfo {author} {\bibfnamefont {T.~R.}\ \bibnamefont
  {Gingrich}}, \bibinfo {author} {\bibfnamefont {G.~M.}\ \bibnamefont
  {Rotskoff}}, \bibinfo {author} {\bibfnamefont {G.~E.}\ \bibnamefont
  {Crooks}}, \ and\ \bibinfo {author} {\bibfnamefont {P.~L.}\ \bibnamefont
  {Geissler}},\ }\href {\doibase 10.1073/pnas.1606273113} {\bibfield  {journal}
  {\bibinfo  {journal} {Proceedings of the National Academy of Sciences}\
  }\textbf {\bibinfo {volume} {113}},\ \bibinfo {pages} {10263} (\bibinfo
  {year} {2016})}\BibitemShut {NoStop}%
\bibitem [{\citenamefont {Ath\`enes}\ and\ \citenamefont
  {Adjanor}(2008)}]{athenes2008measurement}%
  \BibitemOpen
  \bibfield  {author} {\bibinfo {author} {\bibfnamefont {M.}~\bibnamefont
  {Ath\`enes}}\ and\ \bibinfo {author} {\bibfnamefont {G.}~\bibnamefont
  {Adjanor}},\ }\href {\doibase 10.1063/1.2953328} {\bibfield  {journal}
  {\bibinfo  {journal} {The Journal of Chemical Physics}\ }\textbf {\bibinfo
  {volume} {129}},\ \bibinfo {pages} {024116} (\bibinfo {year}
  {2008})}\BibitemShut {NoStop}%
\bibitem [{\citenamefont {Shirts}\ and\ \citenamefont
  {Chodera}(2008)}]{shirts2008statistically}%
  \BibitemOpen
  \bibfield  {author} {\bibinfo {author} {\bibfnamefont {M.~R.}\ \bibnamefont
  {Shirts}}\ and\ \bibinfo {author} {\bibfnamefont {J.~D.}\ \bibnamefont
  {Chodera}},\ }\href {\doibase 10.1063/1.2978177} {\bibfield  {journal}
  {\bibinfo  {journal} {The Journal of Chemical Physics}\ }\textbf {\bibinfo
  {volume} {129}},\ \bibinfo {pages} {124105} (\bibinfo {year}
  {2008})}\BibitemShut {NoStop}%
\bibitem [{\citenamefont {Albaugh}\ and\ \citenamefont
  {Gingrich}(2020)}]{code}%
  \BibitemOpen
  \bibfield  {author} {\bibinfo {author} {\bibfnamefont {A.}~\bibnamefont
  {Albaugh}}\ and\ \bibinfo {author} {\bibfnamefont {T.~R.}\ \bibnamefont
  {Gingrich}},\ }\href {\doibase 10.5281/zenodo.4150012} {\enquote {\bibinfo
  {title} {Estimating reciprocal partition functions to enable design space
  sampling},}\ } (\bibinfo {year} {2020})\BibitemShut {NoStop}%
\bibitem [{\citenamefont {Crooks}\ and\ \citenamefont
  {Chandler}(2001)}]{crooks2001efficient}%
  \BibitemOpen
  \bibfield  {author} {\bibinfo {author} {\bibfnamefont {G.~E.}\ \bibnamefont
  {Crooks}}\ and\ \bibinfo {author} {\bibfnamefont {D.}~\bibnamefont
  {Chandler}},\ }\href {\doibase 10.1103/PhysRevE.64.026109} {\bibfield
  {journal} {\bibinfo  {journal} {Physical Review E}\ }\textbf {\bibinfo
  {volume} {64}},\ \bibinfo {pages} {026109} (\bibinfo {year}
  {2001})}\BibitemShut {NoStop}%
\bibitem [{\citenamefont {Stoltz}(2007)}]{stoltz2007path}%
  \BibitemOpen
  \bibfield  {author} {\bibinfo {author} {\bibfnamefont {G.}~\bibnamefont
  {Stoltz}},\ }\href {\doibase 10.1016/j.jcp.2006.12.006} {\bibfield  {journal}
  {\bibinfo  {journal} {Journal of Computational Physics}\ }\textbf {\bibinfo
  {volume} {225}},\ \bibinfo {pages} {491} (\bibinfo {year}
  {2007})}\BibitemShut {NoStop}%
\end{thebibliography}%

\end{document}